\documentclass[twocolumn,times,trackchanges, dvipsnames]{aastex631}
\usepackage[utf8]{inputenc}
\usepackage{amsmath}

\defcitealias{Carvalho_FUVFUOri_2024ApJ}{C24}

\begin{document}
\title{The Near-Ultraviolet Spectra  of FU Orionis Accretion Disks}

\author{Adolfo S. Carvalho}
\affiliation{Department of Astronomy; California Institute of Technology; Pasadena, CA 91125, USA}
\author{Lynne A. Hillenbrand}
\affiliation{Department of Astronomy; California Institute of Technology; Pasadena, CA 91125, USA}
\author{Gregory J. Herczeg}
\affiliation{Kavli Institute for Astronomy and Astrophysics, Peking University, Beijing 100871, People's Republic of China}
\affiliation{Department of Astronomy, Peking University, Beijing 100871, People's Republic of China}
\author{Kevin France}
\affiliation{Laboratory for Atmospheric and Space Physics, University of Colorado Boulder, Boulder, CO 80303, USA}

\begin{abstract}
    We present the results of the first high-sensitivity NUV (1800 to 3200 \AA) survey of FU Ori objects, using the \textit{Hubble Space Telescope} (HST) STIS spectrograph. We compare new low resolution spectra for 6 sources with predictions from accretion disk models and find that all show emission in excess of the disk model spectrum. The physical properties of the NUV emission excess are very consistent among the sample, with a mean luminosity of $10^{-1.11 \pm 0.4}  \ L_\odot$ and temperature of $16400 \pm 2600$ K -- despite spanning 0.9 dex in $M_*$, 1.3 dex in $\dot{M}$, and 0.7 dex in $L_\mathrm{acc}$. We use the spectra to conclusively rule out the existence of a hot boundary layer in FU Ori accretion disks. We then discuss the source of the excess emission in the context of recent simulations of FU Ori outbursts and boundary layer accretion. The UV spectra also show the often-seen \ion{C}{2}] 2326 \AA\ multiplet and \ion{Mg}{2} 2796/2803 \AA\ doublet, as well as the unusual \ion{Fe}{2}] 2507/2509 \AA\ doublet, a feature that is not seen in the existing UV spectra of other young stellar objects. We measure and compare the luminosities of these lines in outbursting with those in non-outbursting objects.
\end{abstract}

\section{Introduction}\label{sec:introduction}
There is class of young stellar objects (YSOs) called FU Ori objects, or FUOrs \citep{herbig_eruptive_1977}, which are observed to undergo eruptive episodes during which they accrete mass at rates up to 10,000 times greater than typical YSOs \citep{hartmann_fu_1996}. These accretion-driven outbursts can last for several decades, with some appearing to have been in outburst for well over 100 years \citep[e.g., V883 Ori and RNO 54,][]{StromStrom_V883OriDiscovery_1993ApJ, Hillenbrand_RNO54_letter_2023ApJ}. The class is named for FU Ori, which underwent a $\Delta V = 4$ magnitudes outburst in 1937 \citep{Wachmann_FUOri_1954ZA} and has remained at nearly its peak brightness for the past 87 years. Since the start of the FU Ori outburst, approximately 40 FUOrs have been found \citep[][Contreras-Pena et al. 2025, in prep]{connelley_near-infrared_2018}.
The enormous mass accretion rates of FUOrs ($10^{-5} - 10^{-4} \ M_\odot$ yr$^{-1}$) radically alter the geometry of the accretion of material from the disk onto the star. 

In non-outbursting YSOs, the classical T Tauri stars (CTTSs), the accretion geometry is that of magnetospheric accretion \citep{Uchida_MagnetosphericAccretion_1984PASJ}. As material flows toward the star in the inner disk, it eventually meets an opposing pressure from the strong stellar magnetic field, which is stronger nearer the star. There the magnetic field energy density becomes comparable with the ram pressure of the inflowing material and the charged matter is lifted along the stellar magnetic field lines and deposited on the surface of the star \citep[e.g.,][]{Zhu_GlobalSim_2024MNRAS}. The mass flow reaches the star at freefall velocities ($300-400$ km s$^{-1}$) and creates a shock buried in the stellar photosphere. The shock-heated photospheric material at the location of the accretion column emits brightly at ultraviolet (UV) and blue visible wavelengths \citep{Calvet_FunnelFlowStructure_1998ApJ, Hartmann_review_2016ARA&A, 2022AJ....164..201P}. The UV spectra of CTTSs are thus typically dominated by accretion emission, rather than emission from the photosphere of the star.  

In FUOrs, by contrast, when the disk undergoes an instability that dramatically increases the accretion rate, $\dot{M}$, the pressure balance is disrupted. For sufficiently high $\dot{M}$, the magnetosphere is overwhelmed by the accretion flow and the disk engulfs the star. The disk material accretes directly onto the stellar surface modulated by a boundary layer. The boundary layer itself has been theorized to be a region dominated by shear heating as the disk material in Keplerian rotation slows to the rotation rate of the central star \citep{Popham_boundaryLayersInPMSDisks_1993ApJ}. Models of the shear heated boundary layer predict a high maximum temperature, $T_\mathrm{max}$, of the disk at radii approaching the stellar radius, $R_*$ \citep{Popham_boundaryLayerSpectraLineProfiles_1996ApJ}. This high temperature shear-heated component should produce as a bright UV spectrum.

The accretion disk itself can reach temperatures of $7,000 - 8,500$ K \citep{welty_FUOriV1057CygDiskModelAndWinds_1992ApJ,Kenyon_IUE_FUOri_1989ApJ,hartmann_fu_1996, Carvalho_V960MonPhotometry_2023ApJ} and is thus intrinsically UV-bright. The extreme luminosity of the accretion disk and its high maximum temperature have thus far made detection of any UV excess relative to the disk challenging. Identifying a UV excess requires going bluer than 3500-4500 \AA, which is where the hottest components of the disk are brightest. Unfortunately, FUOrs are typically highly extincted \citep{connelley_near-infrared_2018} and appear quite faint in the UV. Altogether, detecting the predicted boundary layer excess would require high sensitivity FUV and NUV spectra.

We recently completed a Hubble Space Telescope (HST) survey aimed at detecting this predicted UV excess. We targeted 6 FUOrs that had previous $U$ or $B$ band measurements bright enough to ensure they would be detected in the NUV, and obtained low resolution spectra spanning $1800-5500$ \AA. We also sought a high resolution, deep spectrum of FU Ori in the FUV, in order to fully characterize the disk spectrum and any potential boundary layer excess emission at the disk-star interface.

In \citet[][hereafter C24]{Carvalho_FUVFUOri_2024ApJ}, we published the finding that the FUV spectrum of FU Ori shows clear excess UV emission but its continuum shape and effective temperature are inconsistent with previous shear-heating models. We found that the continuum was instead more consistent with a magnetic heating or surface accretion shock origin.



In this article we present the UV spectra of the six FUOrs in our survey, including a re-discussion of FU Ori. All six objects have UV emission in excess of the viscous accretion disk. The luminosity of the excess component is correlated with the system luminosity, indicating the UV excess emission mechanism is related to accretion outburst itself. We also discuss the emission lines in the NUV spectra. 

\section{Data} \label{sec:data}
We obtained visible and NUV spectra of 6 FUOrs: FU Ori, V1057 Cyg, V1515 Cyg, V960 Mon, HBC 722, and BBW 76 as part of the Guest Observer (GO) program 17176\footnote{The data can be accessed at the Mikulski Archive for Space Telescopes (MAST) via \dataset[doi: 10.17909/tpk0-8h65]{https://doi.org/10.17909/tpk0-8h65}.}. The sample was selected as the most blue/UV bright FU Ori objects known in 2022 ($B < 16$ mag). The spectra were taken with the Hubble Space Telescope (HST) Space Telescope Imaging Spectrograph (STIS) using the $52^{\prime\prime} \times 2^{\prime\prime}$ arcsec slit in 2 grating settings: G230L (NUV-MAMA) and G430L. An observation log is provided in Table \ref{tab:obs}.

\begin{deluxetable}{ccccc}[!htb]
\caption{HST/STIS Observations Log}\label{tab:obs}
	\tablehead{\colhead{Target} &  \colhead{Date} 	 & \colhead{Obs Time (s)} & \colhead{Grating} &  \colhead{Wavelengths (\AA)} }

\startdata
\hline
FU Ori & 2023-10-08 & 1381 & G230L & $1580 - 3160$ \\
 & 2023-10-08 & 45 & G430L & $2900 - 5700$ \\
\hline
V1057 Cyg & 2023-04-05 & 1886 & G230L & $1580 - 3160$ \\
 & 2023-04-05 & 2818 & G230L & $1580 - 3160$ \\
 & 2023-04-05 & 150 & G430L & $2900 - 5700$ \\
\hline
V1515 Cyg & 2025-01-01 & 1812 & G230L & $1580 - 3160$ \\
 & 2025-01-01 & 2796 & G230L & $1580 - 3160$ \\
 & 2025-01-01 & 150 & G430L & $2900 - 5700$ \\
\hline
HBC 722 & 2023-08-27 & 1856 & G230L & $1580 - 3160$ \\
 & 2023-08-27 & 180 & G430L & $2900 - 5700$ \\
\hline
BBW 76 & 2024-02-15 & 1848 & G230L & $1580 - 3160$ \\
 & 2024-02-15 & 2760 & G230L & $1580 - 3160$ \\
 & 2024-02-15 & 130 & G430L & $2900 - 5700$ \\
\hline
V960 Mon & 2024-09-26 & 130 & G230L & $1580 - 3160$ \\
 & 2024-09-26 & 1810 & G230L & $1580 - 3160$ \\
 & 2024-09-26 & 2722 & G430L & $2900 - 5700$ \\
\enddata
\end{deluxetable}

The spectra cover the wavelength range 1800 \AA\ to 5500 \AA\ at a spectral resolution of $R \equiv \lambda/\Delta\lambda \sim 600$. The signal-to-noise (SNR) ratio per pixel varies depending on the brightness of the source. In order to improve the SNR of the spectra at the bluest wavelengths, we bin the G230L spectra in 30 \AA\ increments, masking out the bright emission features at 2326 \AA, 2505 \AA, and 2800 \AA. The three faintest objects in the sample, V1057 Cyg, BBW 76, and HBC 722, have very little flux to the blue of 2300 \AA, as can be seen in the 2D spectra presented in Appendix \ref{app:2dspec}. 

\begin{figure*}[!htb]
    \centering
    \includegraphics[width=0.99\linewidth]{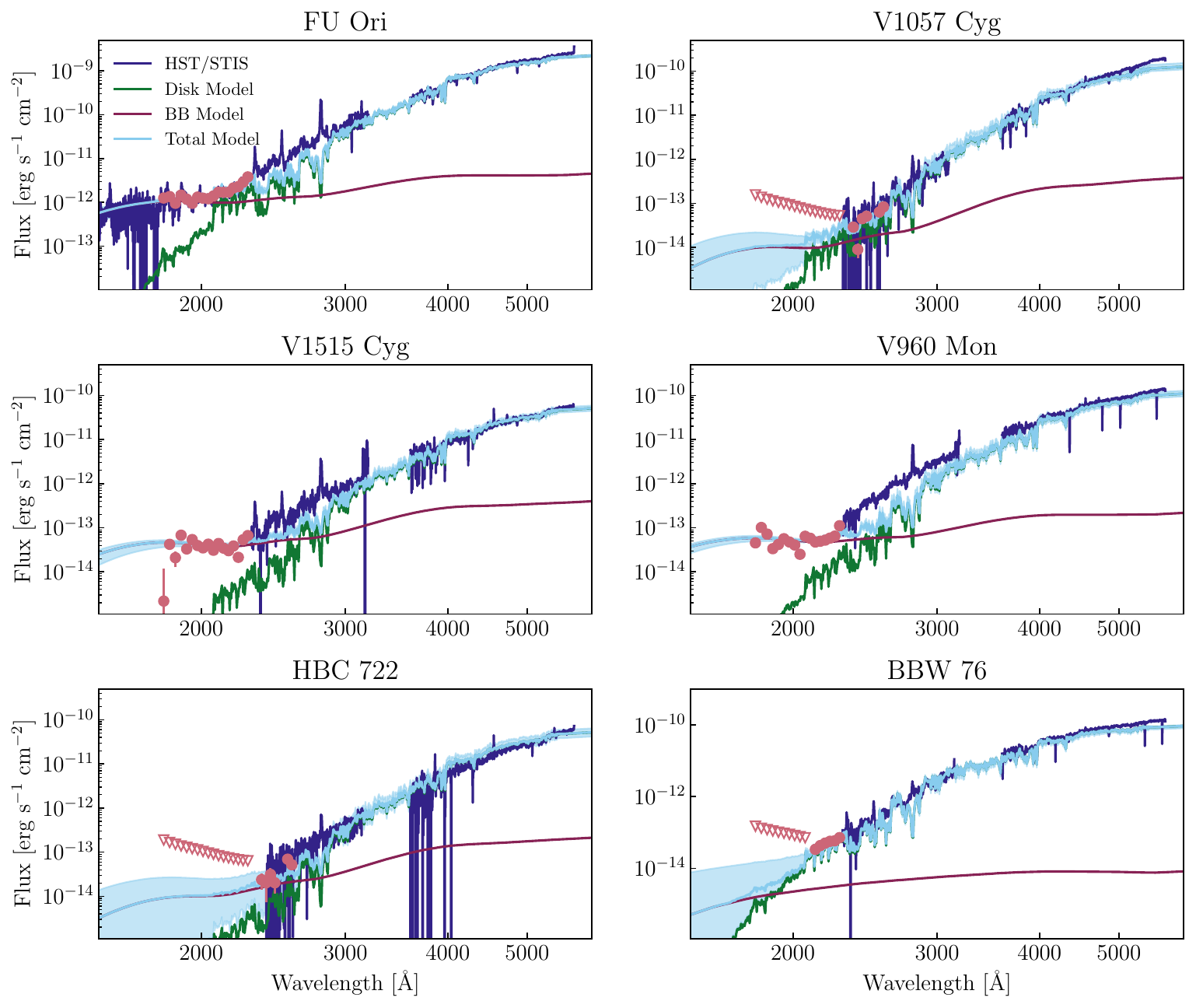}
    \caption{The G230L and G430L spectra for each of our objects (dark blue), along with the disk (green) and disk $+$ UV excess (light blue) models for each of the FUOrs in the survey. The shaded regions show the range of the 16th and 84th percentile models from the MCMC samples. The blackbody component in each panel (maroon) matches the flux level of the binned continuum points (salmon) blueward of 2300 \AA. The flux upper limits for V1057 Cyg, HBC 722, and BBW 76 are shown as empty triangles. 
    }
    \label{fig:DiskModels}
\end{figure*}

For our fainter objects, we consider the source detected if the peak of the spectral trace is $3\times$ the noise value in a given wavelength bin (see Appendix \ref{app:2dspec} for details). V1057 Cyg, BBW 76, and HBC 722 are not detected blueward of 2300 \AA, 2000 \AA, and 2300 \AA, respectively, so for those objects we also computed 3$\sigma$ upper limits on the flux in those wavelength bins. The upper limits are derived from the dark count level, $cts_d$, in each observation\footnote{The average dark counts subtracted per pixel from the image during processing is given in the \texttt{x1d} file header under the header key \texttt{MEANDARK}. We multiply this by the extraction region size, which is saved under the header key \texttt{EXTRSIZE}. This yields the total dark counts contributing to the uncertainty along the extraced object trace. }. We assume that the uncertainty is Poisson-distributed and thus $\sigma_d = \sqrt{cts_d}$. We then converted $\sigma_d$ to an effective flux level using the response function of the detector. The spectra and upper limits are all shown in Figure \ref{fig:DiskModels}.

\section{Modeling the Spectra} \label{sec:diskFits}
We model the NUV spectra with a two component model comprising a thin, viscously heated accretion disk and a single blackbody. We fit the two components generally following the same procedure described in \citetalias{Carvalho_FUVFUOri_2024ApJ}, which we will summarize below. 

The choice of extinction curve cannot be neglected in the analysis of UV data. The assumed shape of the 2175 \AA\ feature can significantly impact continuum shape of dereddened spectra. As in 
\citetalias{Carvalho_FUVFUOri_2024ApJ}, we adopt the extinction curve from \citet{Whittet_ExtinctionCurve_2004ApJ}, which has been demonstrated to be more appropriate for the interstellar environment surrounding YSOs. The extinction curve contains two important features: a weak 2175 \AA\ "bump" and a total-to-selective reddening factor, or $R_V$, of 3.63 \citep[greater than the typically adopted galactic average from][]{cardelli_relationship_1989}. Recent extensive work aimed at measuring $R_V$ along several Milky Way sightlines indicates that $R_V$ values that are greater than the galactic average are found in dense environments near star-forming regions, and that $R_V$ can reach $3.5-3.7$ in the densest regions probed by the study \citep{2025Sci...387.1209Z}. For sources with sufficient SNR at 2100 \AA\ (FU Ori, V1515 Cyg, and V960 Mon, see Figure \ref{fig:DiskModels}), the spectrum does not show the strong 2170 \AA\ absorption feature predicted by the often-used \citet{cardelli_relationship_1989} or \citet{fitzpatrick_unred_1999PASP} extinction laws. 

\subsection{The accretion disk component}

The accretion disk model is the same that we successfully applied to model the spectra of FUOrs and is described in detail in \citet{carvalho_V960MonSpectra_2023ApJ} and \citet{Carvalho_HBC722_2024ApJ}. The primary assumption of the model is that the radial temperature profile of inner disk, $T_\mathrm{eff}(r)$, near the star can be described using the \citet{Shakura_sunyaev_alpha_1973A&A} $\alpha$ disk prescription,
\begin{equation} \label{eq:TProf}
    T^4_\mathrm{eff}(r) = \frac{3 G M_* \dot{M}}{8 \pi \sigma_\mathrm{SB} r^3} \left( 1 - \sqrt{\frac{R_\mathrm{inner}}{r}}  \right)    ,
\end{equation}
where $M_*$ is the stellar mass, $\dot{M}$ is the disk-to-star mass accretion rate, $R_\mathrm{inner}$ is the innermost boundary of the disk, $G$ is the universal gravitational constant, and $\sigma_\mathrm{SB}$ is the Stefan-Boltzmann constant. We modify the temperature profile so that $T_\mathrm{eff}(r < \frac{49}{36}R_\mathrm{inner}) = T_\mathrm{eff}(\frac{49}{36}R_\mathrm{inner}) \equiv T_\mathrm{max}$, following \citep{Kenyon_FUOri_disks_1988ApJ}. This modification has been rigorously tested and upheld as an accurate description of the observed spectrum in both simulations \citep{Zhu_outburst_FUOri_2020MNRAS} and observations in the visible range \citep{Rodriguez_model_2022, Liu_fuorParameterSpace_2022ApJ} and the UV \citepalias{Carvalho_FUVFUOri_2024ApJ}.

\begin{deluxetable*}{c|c|c|c|c|c|c|c|c|c}[!htb]
	\tablecaption{The adopted disk model parameters for each object in the sample. Note that parameters like $\dot{M}$, $R_\mathrm{inner}$ and the emergent $T_\mathrm{max}$ and $L_\mathrm{acc}$ all vary over the course of an outburst. The parameters reported in this Table reflect the state of each system at the time of observation. We show only the uncertainty on $A_V$ here because we incorporated it into our MCMC blackbody fits to account for its large effect on UV luminosity.
 \label{tab:params_disk}}
	\tablewidth{0pt}
	\tablehead{
        \colhead{} & \colhead{$M$} & \colhead{log$\dot{M}$} & \colhead{$R_\mathrm{inner}$} & \colhead{inc} & \colhead{$A_V$} & \colhead{$\sigma(A_V)$} & \colhead{d\tablenotemark{a}} & \colhead{$T_\mathrm{max}$} & \colhead{$L_\mathrm{acc}$\tablenotemark{b}} 
     \\
    	  \colhead{Object} &   \colhead{($M_\odot$)} & \colhead{($M_\odot$ yr$^{-1}$)} & \colhead{($R_\odot$)} & \colhead{(deg)} & \colhead{(mag)} & \colhead{(mag)} & \colhead{(pc)} & \colhead{(K)} & \colhead{($L_\odot$)}
	}
\startdata
\hline
FU Ori & 0.60 & $-4.49$ &  3.52 & 35 & 1.50 & 0.10 & 404 & 5970 & 86 \\
V1057 Cyg & 1.61 & $-5.32$ & 2.67 & 8 & 2.72 & 0.20 &795 & 5858 & 46 \\ 
V1515 Cyg & 1.0 & $-5.68$ & 1.82 & 2 & 2.23 & 0.16 & 960 & 5609 & 89\\ 
V960 Mon & 0.59 & $-4.96$ & 2.69 & 15 & 1.60 & 0.2 & 1120.0 & 5550 & 38 \\ 
HBC 722 & 0.20 & $-4.18$ & 3.65 & 79 & 2.30 & 0.24 & 745.0 & 5840 & 85\\ 
BBW 76 & 0.20 & $-5.38$ & 1.41 & 21 & 0.43 & 0.14 & 1040.0 & 5398 & 9 \\ 
\enddata
\tablenotetext{a}{The distance references are as follows: FU Ori \citep{Kounkel_LamOriDist_2018AJ, Roychowdhury_FUOriV883OriDist_2024RNAAS}; V1057 Cyg \citep{Szabo_V1057cyg_2021ApJ}; V1515 Cyg \citep{Szabo_V1515Cyg_2022ApJ}; V960 Mon \citep{kuhn_comparison_2019}; HBC 722 \citep{kuhn_comparison_2019}; BBW 76 \citep{Gaia_DR3_2023AA}.}
\tablenotetext{b}{This accretion luminosity is \textit{not} the bolometric luminosity of the source.}
\end{deluxetable*}

For each of the objects in the sample, the disk model parameters we use are presented in Table \ref{tab:params_disk}. The disk model component for each object is shown in light blue in Figure \ref{fig:DiskModels}. In the case of FU Ori, we adopted the best-fit parameters we found in \citetalias{Carvalho_FUVFUOri_2024ApJ}. For HBC 722, we adopted the $M_*$ and $R_\mathrm{inner}$ from \citet{Carvalho_HBC722_2024ApJ} and scaled the $\dot{M}$ downward to account for slightly fainter source brightness when the STIS spectrum was taken. For V960 Mon, we followed the prescription from \citet{Carvalho_V960MonPhotometry_2023ApJ} in order to scale the $\dot{M}$ and $R_\mathrm{inner}$ in the model spectrum from the 2017 epoch to the 2023 epoch of the STIS observation. 

The remaining three objects, BBW 76, V1515 Cyg, and V1057 Cyg required new model fits. The details of the fits are given in \citet{Carvalho_Thesis_2026}. We follow the general disk fitting procedure we have applied successfully to the other FUOrs. For these three objects, the free parameters we fit in the disk model are $M_*$, $\dot{M}$, $R_\mathrm{inner}$, $i$, and $A_V$. The SEDs we fit are combinations of the near-infrared (NIR) spectra from \citet{connelley_near-infrared_2018} and the STIS G430L spectra, scaled to the flux of each target in 2015, when the NIR spectra were taken. We also use rotational broadening of lines in high resolution spectra to constrain the maximum Keplerian velocity in the disk \citep[shown in][]{Carvalho_Thesis_2026}. 

\subsection{The excess blackbody component} \label{sec:excessFit}

\begin{figure}[!htb]
    \centering
    \includegraphics[width=0.99\linewidth]{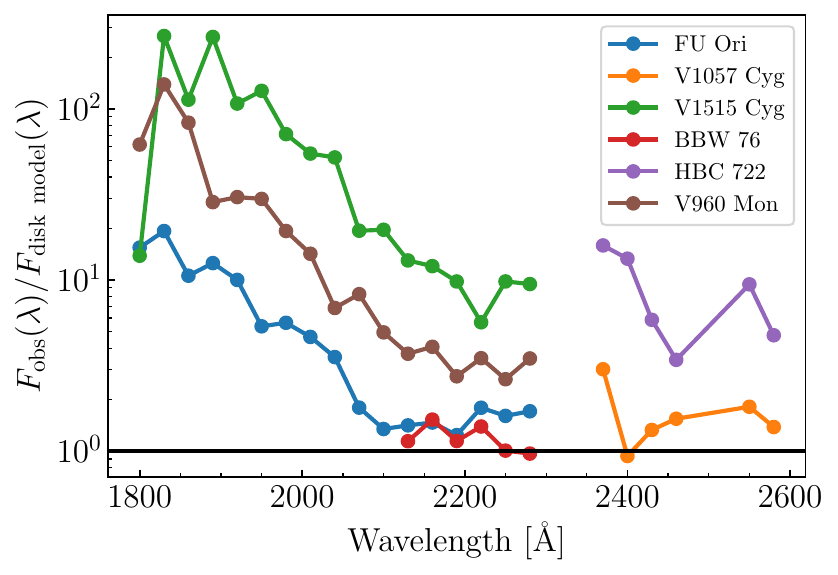}
    \includegraphics[width=0.99\linewidth]{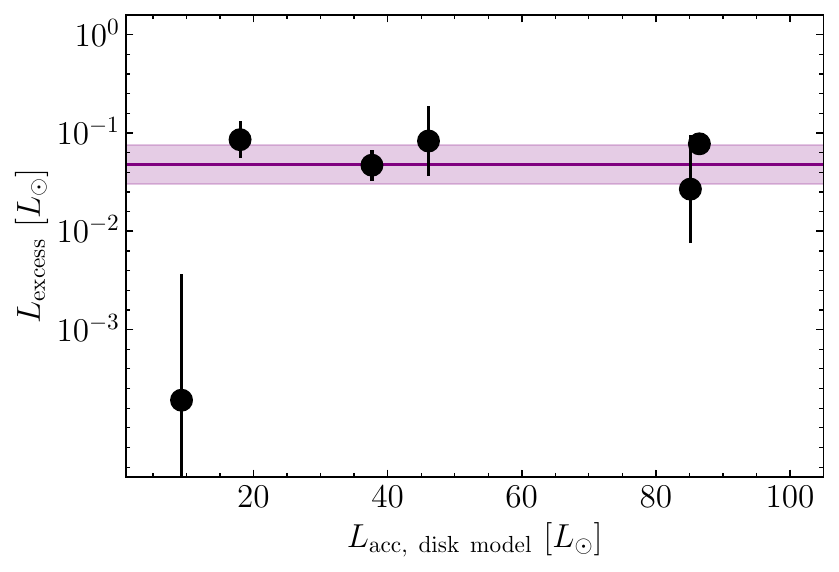}
    
    \caption{ \textbf{Upper Panel:} The ratio of the observed binned STIS spectra used to fit the blackbody excess model and the disk models binned to the same wavelength sampling. Notice that for all the FUOrs in the sample, the NUV flux is at least $2\times$ as bright as the disk model predicts at 2200 \AA\ and for many the flux ratio is $> 10$ (for HBC 722 and V1057 Cyg, this is true at 2400 \AA, despite the lack of data at 2200 \AA). 
    \textbf{Lower Panel:} The $L_\mathrm{excess}$ measurements for each of the FUOrs compared with their disk-model-derived $L_\mathrm{acc}$. The error bars on each point are given by the $1\sigma$ ranges of their posterior distributions. The mean and standard deviation on the UV excess luminosities (excluding BBW 76) are marked by the purple line and shading, respectively.} 
    \label{fig:UVExcesss}
\end{figure}

As can be seen in Figure \ref{fig:DiskModels}, the disk model alone is insufficient to match the NUV continuum emission in all of the sources. The upper panel of Figure \ref{fig:UVExcesss} shows the excess is stronger toward bluer wavelengths, with the observed flux diverging rapidly from the disk model flux for $\lambda < 2200$ \AA. 

In order to model the excess UV emission, we use a single-temperature blackbody, which we showed in \citetalias{Carvalho_FUVFUOri_2024ApJ} accurately matches the FUV and NUV excess emission in FU Ori. The blackbody component model is given by 
\begin{equation}
    F_\lambda = \pi B_\lambda(T_\mathrm{BB})\left( \frac{R_\mathrm{BB}}{d} \right)^2,    
\end{equation}
where $d$ is the distance to the source, $R_\mathrm{BB}$ is the effective radius of the emission region, and $T_\mathrm{BB}$ is the effective temperature of the blackbody, $B_\lambda$. We treat the emission as a disk projected in the plane of the sky, since we do not know its true geometry and to maintain consistency with the assumption in \citetalias{Carvalho_FUVFUOri_2024ApJ}. We assume that the $A_V$ to the UV emission is the same as that of the system, reported in Table \ref{tab:params_disk}. We discuss the validity of this assumption in Section \ref{sec:discussion}. 

For the sources with the highest signal-to-noise spectra (FU Ori, V1515 Cyg, and V960 Mon), we apply the model to the binned continuum points with $\lambda < 2300$ \AA, as this is where the excess relative to the disk model is strongest and where there are few bright emission lines contributing to the continuum bins. For the fainter sources (V1057 Cyg, BBW 76, and HBC 722), we use the bluest wavelength bins in which they are detected, according to the detection criteria in Section \ref{sec:data}. The wavelength ranges used for the fits are reported in Table \ref{tab:params}. 

We then fit the three-parameter model using the log-likelihood based Markov-Chain Monte Carlo nested sampling code \texttt{dynesty} \citep{speagle2020}. The third parameter we vary in our model is $A_V$. This enables us to account for impact of $A_V$ uncertainty on the UV excess luminosity measurements. We sample the $A_V$ values at each iteration from a Gaussian distribution centered on the disk model best-fit and a standard deviation equal to the $\sigma(A_V)$, both of which are given in Table \ref{tab:params_disk} for each source. The median values from the posterior distributions, which we adopt as our best-fits, are reported in Table \ref{tab:params}. For all of the sources, the median $A_V$ in the posterior distributions was within 1$\sigma$ of the injected mean value, as expected, so we do not report these in Table \ref{tab:params}. The posterior distributions themselves are shown in Appendix \ref{app:CornerPlots}.

Due to the lack of signal for $\lambda < 1800$ \AA\ in our spectra, we are unable to adequately constrain the maximum possible temperature of the blackbody. The greatest temperature we allow the walkers to explore is 20000 K, since greater temperatures were ruled out in by the FU Ori FUV spectrum. However, we confidently rule out temperatures below 7,000 K for all of our objects. For FU Ori, V960 Mon, and V1515 Cyg, $T_\mathrm{BB} > 12000$ K is strongly preferred. The posterior distributions follow the expected $R_\mathrm{BB}^2 \propto T_\mathrm{eff}^4$ for a fixed luminosity, enabling us to compute the luminosity of the UV excess as $L_\mathrm{excess} = \pi R_\mathrm{BB}^2 \sigma_\mathrm{SB}T_\mathrm{BB}^4$. We report the luminosities in Table \ref{tab:params} and discuss them in Section \ref{sec:uvExcess}.

\subsection{Results of the blackbody fits and UV excess component properties} \label{sec:uvExcess}

Our spectra of five additional FUOrs confirm the observed properties of the UV excess in FU Ori itself. In particular, the excess components all have temperatures in excess of 12,000 K and filling factors of $< 10^{-3}$ of the stellar surface (Table \ref{tab:params}).

In our sample of 6 objects, the mean $L_\mathrm{excess} = 0.04 \pm 0.02 \ L_\odot$, excluding BBW 76, which shows almost no excess. The mean temperature $T_\mathrm{BB} = 16,200 \pm 2,400$ K and size $R_\mathrm{BB} = 0.05 \pm 0.02 \ R_\odot$ are consistent with what we found for the UV excess of FU Ori using continuum regions of a COS FUV spectrum \citepalias{Carvalho_FUVFUOri_2024ApJ}. 

As can be seen in the lower panel of Figure \ref{fig:UVExcesss}, the $L_\mathrm{excess}$ values show little dependence on the $L_\mathrm{acc}$ of the system. There is also no correlation between $L_\mathrm{excess}$ and $\dot{M}$, $M_*$, $R_\mathrm{inner}$, or $i$. The uniformity of the UV excess across the sample (with BBW 76 as a notable outlier) may have provide hints as to its origin and its relation to boundary layer accretion in general. We discuss the potential emission source in detail in Section \ref{sec:contExcessSource} and the extremely weak excess in BBW 76 in Section \ref{sec:BBW76}.

\begin{deluxetable*}{c|c|c|c|c|c|c|c}[htb]
	\tablecaption{The best-fit blackbody component parameters and NUV emission line luminosities for each object in the sample. 
 \label{tab:params}}
	\tablewidth{0pt}
	\tablehead{
        \colhead{} &  \colhead{Fit Range} & \colhead{log$R_\mathrm{BB}$} & \colhead{$T_\mathrm{BB}$} & \colhead{log$(L_\mathrm{excess})$} & \colhead{$L_\mathrm{MgII}$} & \colhead{$L_\mathrm{CII]}$} & \colhead{$L_\mathrm{FeII]}$} 
     \\
    	  \colhead{Object} &   \colhead{(\AA)} & \colhead{($R_\odot$)} & \colhead{(K)} & \colhead{($L_\odot$)} & \colhead{($10^{-3} \ L_\odot$)} & \colhead{($10^{-3} \ L_\odot$)} & \colhead{($10^{-3} \ L_\odot$)}
	}
\startdata
\hline
FU Ori\tablenotemark{a} & $1800-2300$ & {$-1.3 \pm 0.2$} & {$16,000 \pm 1,000$} & {$-1.06 \pm 0.01$} & {$78.9 \pm 0.3$} & {$7.0 \pm 0.1$} & {$7.0 \pm 0.1$} \\
V1057 Cyg & $2350-2500$ & {$-1.1 \pm 0.4$} & {$13,200 \pm 5,000$} & {$-1.08 \pm 0.35$} & {$26.1 \pm 2.5$} & {$9.9 \pm 0.6$} & {$3.8 \pm 0.2$} \\ 
V1515 Cyg & $1800-2300$ & {$-1.2 \pm 0.3$} & {$16,400 \pm 3,000$} & {$-1.06 \pm 0.18$} & {$31.4 \pm 4.7$} & {$5.3 \pm 0.2$} & {$4.5 \pm 0.7$} \\
V960 Mon & $1800-2300$ & {$-1.4 \pm 0.2$} & {$18,000 \pm 2,000$} & {$-1.32 \pm 0.16$} & {$23.4 \pm 3.5$} & {$1.1 \pm 0.3$} & {$0.9 \pm 0.5$} \\
HBC 722 & $2350-2500$ & {$-1.3 \pm 0.6$} & {$12,200 \pm 5,000$} & {$-1.57 \pm 0.55$} & {$5.20 \pm 0.4$} & {$< 9.0$} & {$0.34 \pm 0.01$} \\
BBW 76 & $2100-2300$ & {$-2.0 \pm 0.7$} & {$8,500 \pm 5,000$} & {$-3.72 \pm 1.28$}  & {$0.38 \pm 0.17$}  & {$0.01 \pm 0.01$}  & {$0.1 \pm 0.01$} \\
\hline
Average\tablenotemark{b} & $\cdots$ & {$-1.28 \pm 0.4$} & {$16,400 \pm 2,600$} & {$-1.11 \pm 0.40$} & $\cdots$ & $\cdots$ & $\cdots$ \\
\enddata
\tablenotetext{a}{Since the excess $T_\mathrm{BB}$ and $R_\mathrm{BB}$ are well constrained for FU Ori in \citetalias{Carvalho_FUVFUOri_2024ApJ}, we adopt those values in this Table.}
\tablenotetext{b}{Excluding BBW 76.}
\end{deluxetable*}

\section{NUV Emission Lines} \label{sec:emissionLines}

The NUV spectra of the FUOrs show 3 prominent emission features: the often-seen \ion{C}{2}] 2326 \AA\ multiplet and \ion{Mg}{2} 2796/2803 \AA\ doublet, as well as the unusual \ion{Fe}{2}] 2507/2509 \AA\ doublet. We discuss each of these features in detail below.

In order to understand the differences between these emission lines in FUOrs and their counterparts in CTTSs, we compute the line luminosities for each feature and discuss the measurements in the following Section. The CTTS we select for comparison is BP Tau. We justify the choice of BP Tau as a representative CTTS in Appendix \ref{app:BPTau}. We measure its line luminosities in the STIS G230L spectrum obtained in 2002 in the HST GO program 9081 \citep[PI: N. Calvet;][]{Kravtsova_BPTauUVSpectrum_2003AstL}, which was accessed through MAST. We adopt a distance of 131 pc \citep{Gaia_DR3_2023AA} and an $A_V = 0.41$ mag to the source \citep{herczeg_survey_2014}.

\begin{figure}[!htb]
    \centering
    \includegraphics[width=0.98\linewidth]{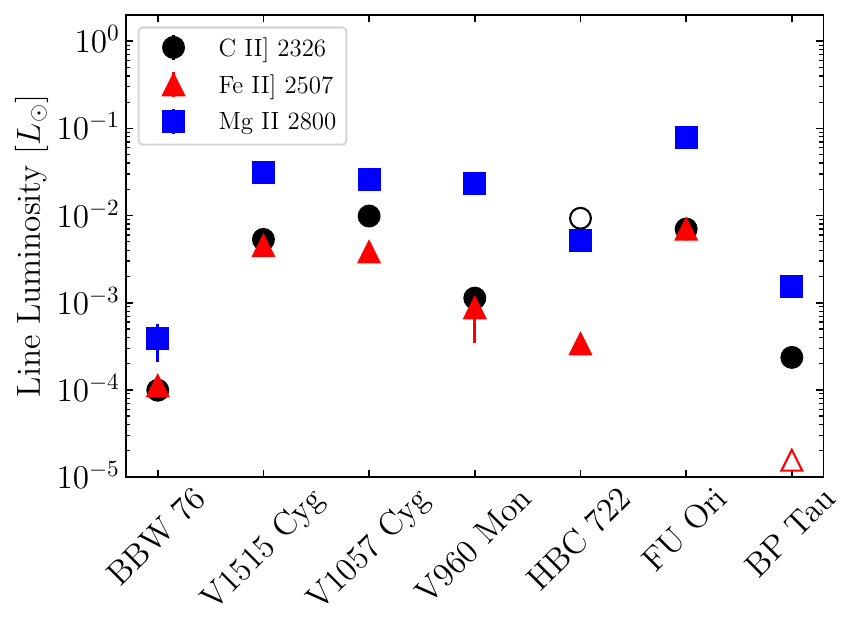}
    \includegraphics[width=0.98\linewidth]{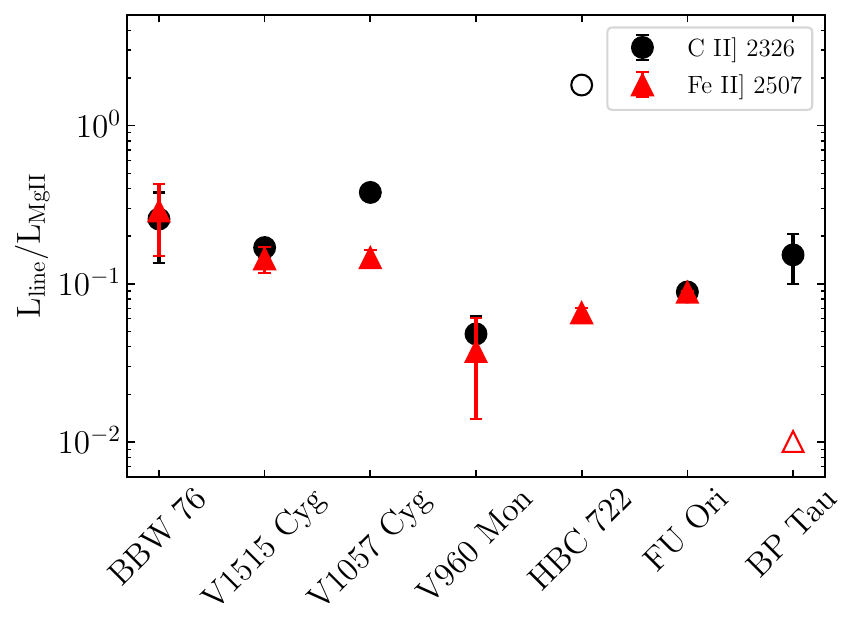}
    \includegraphics[width=0.98\linewidth]{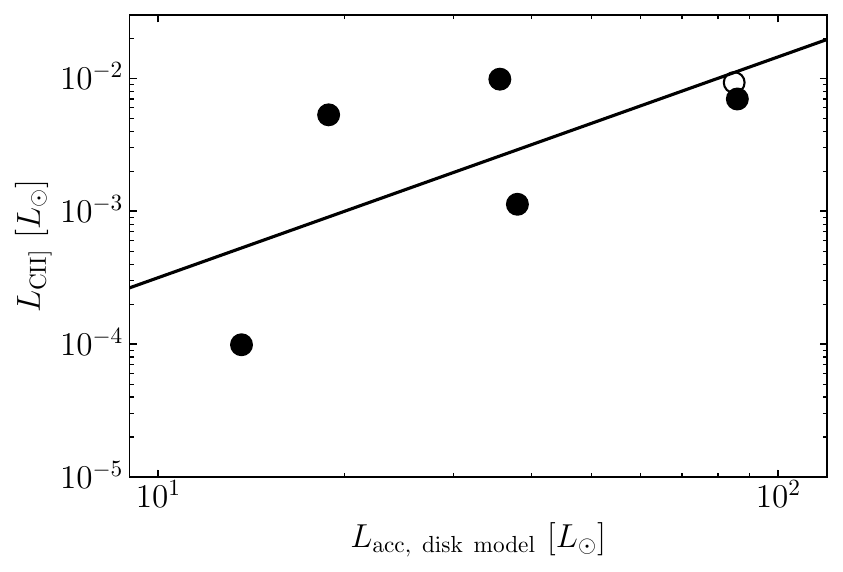}
    \caption{\textbf{Top:} The \ion{C}{2}] (black circles), \ion{Fe}{2}] (red triangles), and \ion{Mg}{2} (blue squares) line luminosities for the FUOr sample and the CTTS BP Tau. Error bars are drawn but are smaller than the symbols for most measurements. The empty circle (triangle) for HBC 722 (BP Tau) represents the 3$\sigma$ upper limit on the \ion{C}{2}] (\ion{Fe}{2}]) luminosity. The FUOrs are sorted left-to-right in order of increasing $L_\mathrm{acc}$. \textbf{Middle:} The same line \ion{C}{2}] and \ion{Fe}{2}] luminosities divided by the \ion{Mg}{2} luminosities. \textbf{Bottom:} The $L_\mathrm{CII]}$ measurements plotted against the $L_\mathrm{acc}$ for the FUOrs. The black line shows the best-fit relation described in Section \ref{sec:CII}.}
    \label{fig:LineLuminosities}
\end{figure}

The line fluxes were measured from the dereddened spectra, which were extinction-corrected using the $A_V$ values reported in Table \ref{tab:params_disk} and the \citet{Whittet_ExtinctionCurve_2004ApJ} extinction law. 
We then fit a pseudocontinuum near the line using the $\mathtt{specutils}$ function $\mathtt{fit\_continuum}$ and compute the line flux via direct integration. The integration limits for each lines are: $2310-2340$ \AA\ (\ion{C}{2}]), $2503-2515$ \AA\ (\ion{Fe}{2}]), and $2780-2820$ \AA\ (\ion{Mg}{2}). The flux is then converted to a luminosity using the source distances in Table \ref{tab:params}. The luminosity measurements are shown in Figure \ref{fig:LineLuminosities}. 

\begin{figure}[htb]
    \centering
    \includegraphics[width=0.99\linewidth]{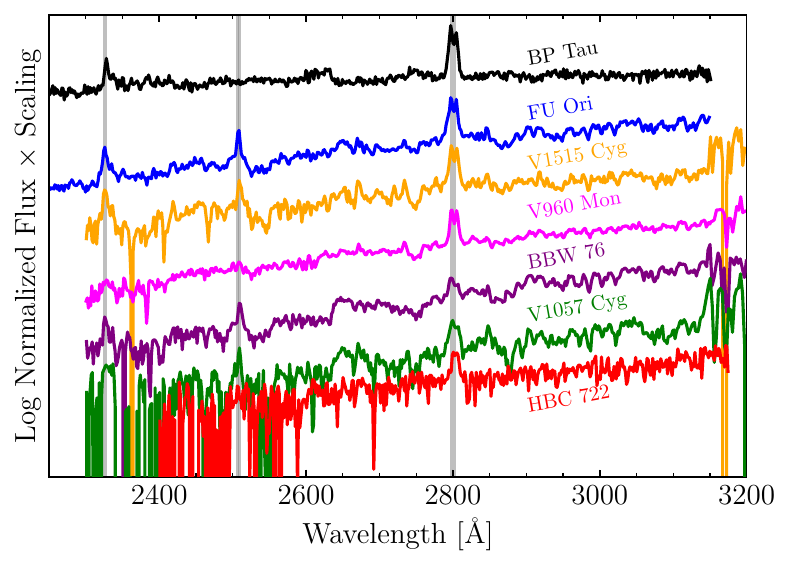}
    \caption{The STIS G230L spectra of the 6 FUOrs and the CTTS BP Tau for reference. The spectra are sorted top-to-bottom in order of decreasing signal-to-noise. The grey vertical lines mark the locations of the emission features discussed in Section \ref{sec:emissionLines}. }
    \label{fig:NUVStacked}
\end{figure}

\subsection{The Mg II doublet}
The brightest feature in the NUV spectra of all 6 objects is the Mg II 2796/2803 \AA\ doublet. Although doublet luminosity has been shown to correlate with the accretion luminosity of CTTSs, it can display P Cygni profiles typical of lines that trace outflows and has been seen in non-accreting, but magnetically active, YSO spectra \citep{Ingleby_NUVEmission_CII_2013ApJ, Xu_CII_winds_CTTSs_2021ApJ}. The high resolution 2001 HST/STIS E230M spectrum of FU Ori also reveals strong outflow absorption in the Mg II doublet, as both features are in clear P Cygni profiles \citep{Kravtsova_FUOriSTIS_2007AstL}. 

Despite the unknown degree of outflow absorption in the profiles, we measure and report the doublet luminosity for each object. The high resolution E230M spectrum of the FU Ori \citep{Kravtsova_FUOriSTIS_2007AstL} shows that the deep blue-shifted absorption removes approximately half of the flux from the total line emission. Therefore, our reported luminosities should be considered lower limits on the intrinsic \ion{Mg}{2} doublet luminosity of FUOrs. 

Visually, the \ion{Mg}{2} line strengths look similar across the sample, with the exception of BBW 76 and V1057 Cyg. In these two objects, the \ion{Mg}{2} emission is much weaker relative to the other two emission features, an effect that is quantified in the ratios plotted in Figure \ref{fig:LineLuminosities}. We discuss the potential implications of this in Section \ref{sec:discussion}.

\subsection{The C II] complex at 2326 \AA} \label{sec:CII}
The \ion{C}{2}] emission feature at 2326 \AA\ is in fact a blend of multiplets from three species: \ion{C}{2}], \ion{Si}{2}], and \ion{Fe}{2}]. The features in the wavelength range over which we measure line flux that contribute most strongly to the line are: \ion{C}{2}] 2324.21, 2325.4, 2326.11, 2327.64, and 2328.83, \ion{Si}{2}] 2329.23, and 2335.12/2335.32, and \ion{Fe}{2}] 2328.11. In the STIS E230M spectrum of FU Ori, the \ion{Si}{2}] 2335.12/2335.32 doublet has a blue-shifted absorption/red-shifted emission P Cygni profile. The emission from the \ion{Fe}{2}] features in the region is also mildly red-shifted, suggesting they may be in P Cygni profiles as well. This indicates that at least the \ion{Fe}{2}] and \ion{Si}{2}] features trace an outflow in the system. 

Looking closely at the \ion{C}{2}] multiplet in the STIS E230M spectrum of FU Ori reveals that the lines are unlikely to arise from an optically thin plasma in collisional equilibrium. In CTTSs, this assumption has enabled the estimation of temperature and number density of the line-emitting material based on flux ratios of lines in this complex  \citep{LopezMartinez_FUOriCII_2014MNRAS}. We test this assumption in FU Ori using line emissivities from the CHIANTI database. Over a broad range of electron densities ($n_e = 10^2 - 10^{15}$ cm$^{-3}$) and temperatures ($T_e = 10^4 - 10^{6}$ cm$^{-3}$), the ratio of \ion{C}{2}] 2325.4/2326.11 intensities is at most $0.2-0.3$. In the FU Ori E230M spectrum, the ratio is 0.74, indicating the level population of the multiplet is in disequilibrium. 

Although in CTTSs the \ion{C}{2}] 2326 \AA\ luminosity is a reliable tracer and diagnostic of accretion, its luminosity in FUOrs does not follow the same scaling. If we apply the \citet{Ingleby_NUVEmission_CII_2013ApJ} relation found for CTTSs, $\log L_\mathrm{acc} = 1.1 \log(L_\mathrm{CII]}) + 2.7$, the $L_\mathrm{acc}$ values for the FUOrs are underpredicted by factors of $40-160$. There is some mild correlation between the $L_\mathrm{CII]}$ and $L_\mathrm{acc}$ values, however, which can be fit by $\log(L_\mathrm{acc}) = 1.66 \log(L_\mathrm{CII]}) - 5.16$ (omitting the HBC 722 upper bound in the fit). The best fit is shown in Figure \ref{fig:LineLuminosities}.

\subsection{The Fe II doublet and its origin}
There is a strong feature at 2508 that we identified as the \ion{Fe}{2}] 2507/2509 doublet in the spectrum of FU Ori \citepalias{Carvalho_FUVFUOri_2024ApJ}. The luminosity of the line closely matches the luminosity of the \ion{C}{2}] feature in all of the sources, even matching the $L_\mathrm{CII]}$ upper limit in HBC 722. This indicates that the feature may also correlate with the $L_\mathrm{acc}$ of FUOrs (except HBC 722). However, as we discuss below, the origin of the feature is not well-understood. 

The presence of the \ion{Fe}{2}] 2507/2509 doublet in the spectra of FUOrs is surprising. It is not seen at all in the spectra of other accreting YSOs, but has been observed in spectra of evolved systems like symbiotic binaries, $\eta$ Carinae, and chromospherically-active giants \citep{Johansson_FluorescenceLines_1993PhST, Reza_FeII2507InGiants_2025A&A}, indicating it may be diagnostic of boundary layer accretion. The feature is not expected to appear as brightly as we see it in the FUOr spectra in isolation. For an optically thin plasma in local thermodynamic equilibrium, there are several \ion{Fe}{2} emission lines around $2495-2520$ \AA\ that should be as bright as or brighter than the 2507/2509 doublet \citep{Dere_CHIANTI_1997A&AS,delZanna_chiantiXVI_2021ApJ}. However, the high resolution STIS E230M spectrum of FU Ori confirms that the doublet is much brighter than any nearby \ion{Fe}{2} features \citep{Kravtsova_FUOriSTIS_2007AstL}. 

The strength of the doublet relative to other \ion{Fe}{2} features in the same wavelength range is attributed to fluorescence via one of two proposed mechanisms: photoexcitation by accidental resonance or photoexcitation by continuum resonance \citep[termed PAR or PCR, respectively, ][]{Johansson_FluorescenceLines_1993PhST}. Both mechanisms rely on photons in the 1000-1300 \AA\ wavelength range to pump \ion{Fe}{2} transitions that populate energy levels around 11-12 eV, which eventually cascade downward and through the 2507/2509 lines \citep{Johansson_FluorescenceLines_1993PhST}\footnote{The specific transitions are identified by \citet{Johansson_FluorescenceLines_1993PhST} as 5p$^6$F$_{9/2}^\circ$ $\rightarrow$ 4s c$^4$F$_{7/2}$ and 4p$^4$G$_{9/2}^\circ$ $\rightarrow$ 4s c$^4$F$_{7/2}$ for 2507 \AA\ and 2509 \AA, respectively. }. While it is beyond the scope of this paper to model the emission of this doublet in detail, we did investigate the PAR/PCR mechanisms discussed in \citet{Johansson_FluorescenceLines_1993PhST} and their plausibility in FUOrs.

The PAR mechanism relies on the coincidence between the wavelength of an existing emission line and that of a transition in another species. In the case of the 2507/2509 doublet, \citet{Johansson_FluorescenceLines_1993PhST} propose that very red-shifted Ly$\alpha$ might excite electrons via transitions at 1217.85 \AA\ and 1218.21 \AA, which would require Ly$\alpha$ emission reaching $500-600$ km s$^{-1}$. Alternatively, $-70$ km s$^{-1}$ emission by the \ion{O}{5} 1218.5 \AA\ line could also excite the same transition. In \citetalias{Carvalho_FUVFUOri_2024ApJ}, we found high velocity red-shifted Ly$\alpha$ in the high resolution FUV spectrum of FU Ori, as well as blue-shifted emission from \ion{O}{5} at 1371 \AA\ with a velocity of $-80$ km s$^{-1}$. Though this suggests Ly$\alpha$ emission as a promising PAR source, CTTSs have extremely bright Ly$\alpha$ at velocities of $500-600$ km s$^{-1}$ \citep{arulanantham_lyAlpha_2023ApJ} but do not show the Fe II doublet. 

In the case of PCR, all that is necessary is bright continuum emission in the correct wavelength range to excite the desired transition. \citet{Johansson_FluorescenceLines_1993PhST} propose that \ion{Fe}{2} ground transitions around 1100-1115 \AA\ can be excited by continuum emission. A clear signature of this excitation mechanism would be a series of absorption features in 1100-1115 \AA\ wavelength range, though that may depend on the geometry of the absorbing material. There is also the case of the 1785 \AA\ \ion{Fe}{2} feature, which can result from PCR of ground transitions around 1270 \AA\ \citep{HempeReimers_FeII1785_1982A&A}. While the signal-to-noise ratio of the FUV spectrum of FU Ori is insufficient to detect the predicted line absorption, PCR would explain the strength of the bright feature at 1785 \AA\ seen in the FUV spectrum of FU Ori \citetalias{Carvalho_FUVFUOri_2024ApJ}. Given the bright FUV continuum of FU Ori, which we showed extends beyond 1150 \AA\ \citepalias{Carvalho_FUVFUOri_2024ApJ}, it is possible that PCR in the 1100 \AA\ range could pump the upper transitions of the 2507/2509 doublet. 

Both PAR and PCR can also produce fluorescent line emission at wavelengths around 9000 \AA, but the lines would be significantly fainter than the accretion disk and therefore not observable in the spectra of FUOrs. Clearly identifying or ruling out PAR/PCR as the source of the transition pumping photons would require both careful modeling of this mechanism in FUOrs and a much more sensitive FUV spectrum of FU Ori.

\section{Discussion} \label{sec:discussion}

In Section \ref{sec:uvExcess}, we demonstrated that all six FUOrs in the sample possess excess UV continuum emission that reaches more than $10\times$ the disk model flux, as shown in Figure \ref{fig:UVExcesss}. We also found that the excess emission spans a narrow range of luminosities ($L_\mathrm{excess} \sim 0.04 \ L_\odot$), temperatures ($T_\mathrm{BB} \sim 12000-18000$ K), and physical size ($R_\mathrm{BB} \sim 0.02-0.06 \ R_\odot$). Despite the 0.9 dex range in stellar mass and 1.2 dex range in mass accretion rate, the $L_\mathrm{excess}$ excess varies little across the sample. Here, we discuss the potential source of the emission in the context of the classical boundary layer accretion model for FUOrs and the exceptional case of the source BBW 76.

\subsection{The UV Excess Emission in FUOrs compared with CTTSs} \label{sec:comparison}

To contextualize the luminosity of the UV excess in FU Ori objects relative to other YSOs, we compare our measurements with the UV survey of CTTSs compiled by \citet{yang_HST_TTS_FUV_Survey_2012ApJ}. We restrict our comparison to the lower mass and lower luminosity subset of CTTSs ($L_\mathrm{acc} < 1 \ L_\odot$), since they better reflect the pre-outburst FUOrs in our survey.

Directly comparing the UV emission from both sets of objects is challenging due to the different accretion geometries in each. In CTTSs, for instance, a large fraction ($10-50 \%$) of the accretion luminosity is emitted at $2000 < \lambda < 3100$ \AA\ \citep[termed $L_\mathrm{space}$ in][]{2025arXiv250701162P}. Integrating our disk $+$ blackbody models in the range $2000 < \lambda < 3100$ \AA, we find a median $L_\mathrm{space} = 10^{-1.7 \pm 0.4} \ L_\odot$, which is only a bit smaller than $L_\mathrm{excess}$. Dividing by the $L_\mathrm{acc}$ values in Table \ref{tab:params_disk}, we obtain $L_\mathrm{space}/L_\mathrm{acc} = 10^{-3.4 \pm 0.2}$. In other words, only a very small fraction of the total $L_\mathrm{acc}$ in FUOrs is emitted in the NUV, though the actual NUV luminosities themselves are high.

Comparing the luminosities of the CTTSs and FUOrs in the FUV is further complicated by the lack of FUV spectra in our survey (with the exception of FU Ori). Although we do not have FUV spectra for the other 5 objects in the sample, we can extrapolate the blackbodies into the FUV. The blackbodies have similar best-fit $T_\mathrm{BB}$ and $R_\mathrm{BB}$ to the FUV-derived best-fit values for FU Ori, indicating the FUV continuum emission in our sample is likely well-represented by these blackbody fits. 

We then integrate the blackbodies in the range $1250-1700$ \AA, as was done to measure the $L_\mathrm{FUV}$ of CTTSs reported in \citet{yang_HST_TTS_FUV_Survey_2012ApJ}. The mean $L_\mathrm{FUV, FUOr} = 0.012 \pm 0.007 \ L_\odot$ following this process. This is $7.5\times$ greater than the mean $L_\mathrm{FUV,CTTS} = 0.0016 \pm 0.002 L_\odot$ for the CTTSs in the \citet{yang_HST_TTS_FUV_Survey_2012ApJ} with $L_\mathrm{acc} < 1 \ L_\sun$. The true difference between the FUV continuum luminosities of the two samples is likely greater, since the FUV spectra of CTTSs are emission-line-dominated, with continuum emission matching the typical accretion shock models for $T_\mathrm{eff} = 10^4$ K \citep{France_FUVContinuum_2011ApJ}.

\subsection{The Missing Boundary Layer in FUOrs}\label{sec:BL}

In the viscous accretion disk model applied to FUOrs \citep{1974MNRAS.168..603L, Kenyon_FUOri_disks_1988ApJ}, only half of the viscous dissipation of gravitational energy ($GM_*\dot{M}/R_\mathrm{inner}$) is expected to be radiated by the disk itself. A key assumption is that as material approaches the star, it must decelerate from its large Keplerian rotation rate to the much slower stellar rotation rate. This requires dissipating the remaining half of the initial gravitational energy over a region known as the dynamical boundary layer.

If the energy dissipation mechanism in the dynamical boundary layer is viscosity, then the energy should be efficiently radiated away as the material is heated. The dynamical boundary layer would then have a luminosity as great as the accretion disk due to this extreme shear ($L_\mathrm{BL} = L_\mathrm{acc})$. In our survey, the median value of $L_\mathrm{excess}/L_\mathrm{acc}$ measured for the 6 FUOrs is $10^{-2.9 \pm 0.4}$, which is several orders of magnitude below what would be expected of a shear-heated boundary layer. 

A potential solution is to assume that the boundary layer may be confined to latitudes near disk midplane and obscured from our view due to the surface accretion layer. Could the UV spectrum be reproduced by adopting a higher temperature \citep[$T_\mathrm{BL}\sim30,000$ K,][]{hartmannKenyon_FUOrs_1985ApJ} and luminosity ($L_\mathrm{BL} = 85 \ L_\odot$) for the boundary layer but obscuring it under some larger $A_V$? We explored this using the FUV spectrum of FU Ori and found that even for large $T_\mathrm{BL}$, adopting $A_V > 2$ mag results in model spectra that underpredict the flux for $\lambda < 1600$ \AA. 

We note that the obscuring material is probably not dust, and thus $A_V$ may not be an appropriate proxy for its absorption. However, if we were to instead model the gas absorption against the 30,000 K continuum, the absorption would be dominated by the bound-free transitions in metallic species like C and Si, which have even greater continuous opacities than dust at FUV wavelengths \citep{TravisMatsushima_Opacities_1968ApJ}. We do not see these features in the FUV spectrum of FU Ori, but it is possible that the continuum sensitivity is insufficient to detect them. Perhaps future large UV/optical telescopes with greater senstivity will reveal these continuum features in the FUV spectra of FUOrs.  In either case, for the large column densities necessary to diminish the observed $L_\mathrm{BL}/L_\mathrm{acc}$ by several orders of magnitude, the FUV continuum at $\lambda < 1600$ \AA\ is severely underestimated.

The upper limit on extinction also places an upper limit of $0.6 \ L_\odot$ on the luminosity of an obscured boundary layer. It is therefore unlikely that the FUV continuum in FU Ori is due to a classical boundary layer obscured by disk material along the line of sight. Furthermore, \citep{Popham_boundaryLayersInPMSDisks_1993ApJ} found that for the high accretion rates in FUOrs, the boundary layer should extend over as much as $0.2 \ R_*$, which is incompatible with the median $R_\mathrm{BB} = 0.05 \ R_\odot$.

The uniformly low luminosity and small spatial scale of the UV excess in FUOrs is inconsistent with the classical models for a hot boundary layer. In \citetalias{Carvalho_FUVFUOri_2024ApJ}, we also provide a more extensive discussion of other models for shear-heated boundary layers, both in the literature of YSOs and compact objects, and why the UV excess in FU Ori is inconsistent with those. 

In short, the remaining gravitational energy of the gas near the stellar surface is not dissipated by heating the boundary layer. It is instead likely that the pressure support from the strong magnetic fields near the surface of the star slows the material as it approaches. The interacting disk and stellar fields in the boundary layer are capable of efficiently removing large amounts of angular momentum via Maxwell stresses \citep{Takasao_BoundaryLayer_2025ApJ}. The magnetic field is thus capable of slowing material down to the stellar rotation rate without strongly heating the gas.

\subsection{The source of the UV continuum excess in FUOrs} \label{sec:contExcessSource}

As we established in Section \ref{sec:BL}, the UV excess emission in FUOrs does not arise from a classical hot boundary layer. What could the source of the excess be, then?

In CTTSs, the UV excess emission arises from the heated pre-shock and post-shock material in the photosphere \citep{Calvet_FunnelFlowStructure_1998ApJ, Hartmann_review_2016ARA&A}. This material is expected to be $8,000-10,000$ K, while the shock temperature is expected to be $\sim 9 \times 10^5$ K due to the high free-fall velocities of matter flowing along the accretion funnel. 


The picture we presented in \citetalias{Carvalho_FUVFUOri_2024ApJ} was that the FUV continuum in FU Ori is from a directly-observed accretion shock that is much cooler than that of CTTS systems. The velocity of material producing this shock is the surface accretion flow velocity at the disk-star interface, which is found to be $\sim 40 \%$ of the Keplerian velocity near the star \citep{Zhu_outburst_FUOri_2020MNRAS}. For most of our sources, the Keplerian velocity at $r= R_*$ is $160 \pm 20$ km s$^{-1}$, indicating a surface flow velocity of $v_\mathrm{surf} = 64 \pm 8$ km s$^{-1}$. If we assume that the resulting shock is a strong shock, we estimate the shock temperature: $T_s = \frac{3}{16} \frac{\mu m_H}{k}v_\mathrm{surf}^2 = 45,500$ K, where $k$ is the Boltzmann constant, $m_H$ the mass of the hydrogen atom, and $\mu$ the mean atomic weight, set to 0.5 here since the material is likely fully ionized. $T_s = 45,500$ is much greater than the maximum $T_\mathrm{BB}$ in the sample.

If we assume that the the magnetic pressure support from the star/disk field at the boundary further slows the material \citep{Takasao_BoundaryLayer_2025ApJ}, we can adopt 40 km s$^{-1}$ as the inflow shock velocity and compute $T_s = 17,300$ K, which is in better agreement with the $T_\mathrm{BB}$ values. The range of temperatures observed in the sample can then be reproduced by increasing or decreasing $v_\mathrm{surf}$ by just $5$ km s$^{-1}$, while the range of $v_\mathrm{Kep}(R_*)$ in our sample is more than $100$ km s$^{-1}$. We observe a weak correlation between $T_\mathrm{BB}$ and $v_\mathrm{Kep}(R_*)$, supporting the idea that $v_s \propto v_\mathrm{Kep}(R_*)$ subject to some braking effect. 

It is also conceivable that the magnetic deceleration of the surface flow may be more efficient in systems with greater $v_\mathrm{Kep}(R_*)$. The magnetic deceleration depends in part on the strength of the bundled fields near the stellar surface, which depend on the shear between the star and disk \citep{Takasao_BoundaryLayer_2025ApJ}. Systems with larger $v_\mathrm{Kep}(R_*)$ may thus have stronger fields at $R_*$ and more effectively brake the more rapid flow. In this way the deceleration may regulate the accretion flow velocity to produce the narrow range of observed temperatures. 

An alternative hypothesis is that the UV emission arises from material above the star that is heated by rapidly recurring magnetic reconnection events at the star-disk interface. These events are seen in simulations of boundary layer accretion and are proposed to be the energy source heating the material emitting soft X-rays in FUOrs \citep{Takasao_BoundaryLayer_2025ApJ}. This would also explain the high temperatures of the X-rays in FUOrs \citep{kuhn_comparison_2019}, as well as the hour-to-hour variability of X-ray luminosity in FU Ori \citep{Skinner_FUOri_Xray_2010ApJ}.The UV emission may then be from material between the stellar surface and the hot corona above. 

\subsection{The strange case of BBW 76} \label{sec:BBW76}

A potential hint to the UV excess emission source lies in the outlier of our sample: BBW 76. We do detect the UV excess in the source, but it has an $L_\mathrm{excess}$ that is 13$\times$ smaller than the median in the sample. Our upper limits on the flux blueward of 2000 \AA\ also rule out the potential of a hotter or brighter component with $L_\mathrm{excess} > 0.02 \ L_\odot$ (5$\times$ fainter than the rest of the sample), despite an accretion luminosity of $L_\mathrm{acc} = 9 \ L_\odot$. Furthermore, BBW 76 has an NUV line emission spectrum that is distinct from the other FUOrs. The \ion{Mg}{2} 2800 \AA\ doublet is much less luminous than in the other objects, even the CTTS BP Tau, and there is a strong continuum break at 2650 \AA, which appears only weakly in the other objects.  

While its large (relative to CTTSs) accretion luminosity confirms it is still in outburst, recent photometry from ASAS-SN and NEOWISE \citep{2014ApJ...788...48S,Mainzer_neowise_2011ApJ,2023arXiv230403791H} shows that BBW 76 has faded monotonically by more than $\Delta V = 0.3$ mag \citep[and more than 0.7 mag since 2002,][]{2020A&A...644A.135S} and $\Delta W1 = 0.8$ mag in 10 years, while a recent near-infrared spectrum of the source shows significant spectroscopic evolution in the past 5 years (private communication). It is possible that the source has entered an intermediate stage between boundary layer accretion and magnetospheric accretion, wherein the proximity of the inner disk to the central star still produces a bright disk-dominated visible range continuum. 

A transition between boundary layer accretion and magnetospheric accretion could result in a rapid decline in the UV excess luminosity. In the shock case, once the magnetospheric funnel flows re-establish themselves, the disk and stellar fields will separate spatially, lowering the magnetic flux (and therefore magnetic pressure) near the location of the former boundary layer. The pressure support will no longer slow the surface accretion flow, enabling it to impact the photosphere at velocities more similar to the CTTS case, embedding the shock in the photosphere. The lower $T_\mathrm{BB}$ in BBW 76 implies we may be seeing the heated photosphere re-processing this buried shock, rather than the shock itself. 

The magnetic reconnection events that may produce the excess rely on the interaction of the stellar and disk fields at the boundary layer \citep{Takasao_BoundaryLayer_2025ApJ}. Once again, as the magnetospheric accretion funnels re-establish themselves, the disk and stellar surface will be separated, lessening the frequency of reconnection events. The heating would diminish and the temperature and luminosity of the emission would decrease as the system evolves away from boundary layer accretion.

\section{Conclusions}
Our recent HST/STIS survey of FUOrs reveals that FUV continuum emission in excess of the viscous accretion disk is common. The typical properties of the FUV continuum in the sample are: $L_\mathrm{excess} = 10^{-1.11 \pm 0.4} \ L_\odot$, $T_\mathrm{BB} = 16,400 \pm 2,600$ K, and $R_\mathrm{BB} = 0.05 \pm 0.04 \ R_\odot$. The consistent luminosities, temperatures, and physical sizes of the emitting material indicate a common source for the emission in all FUOrs. 

The NUV line emission from the sample is also bright, though not as bright as predicted by correlations between line and accretion luminosity in CTTSs. The presence of strong \ion{Fe}{2}] 2507/2509 doublet emission relative to other \ion{Fe}{2} features also imply that the physical conditions of the plasma emitting the NUV lines differ significantly from those in the NUV emitting accretion flows of CTTSs.

Recent magnetohydrodynamical simulations of both the FU Ori system and boundary layer accretion in protostars support two potential origins for the UV continuum emission: a shock where the surface accretion flow impacts the stellar surface or material above the star heated by magnetic reconnection events at the star/disk interface. The comparison of the properties of the UV emission and physical properties ($M_*$, $\dot{M}$, $R_\mathrm{inner}$) of the sample do not highlight a preference for one mechanism over another. 

More detailed study and modeling of the emission lines in the NUV spectra will help to establish whether they share a common origin with the UV continuum emission. A UV-sensitive spectrograph on a future large space telescope will be critical for expanding the number of FUOrs with NUV and FUV spectroscopy beyond the 6 covered in this study. 


\software{\texttt{Astropy} \citep{astropy_2013,astropy_2018,astropy_2022}, \texttt{NumPy} \citep{harris2020array}}


\bibliography{references}{}
\bibliographystyle{aasjournal}

\appendix 

\restartappendixnumbering

\section{STIS 2D Spectra and Detection Criteria For Fainter Sources} \label{app:2dspec}

The 2 dimensional recitified \texttt{x2d} G230L spectra of each object in the sample. For sources with 2 G230L exposures (see Table \ref{tab:obs}), the images have been coadded via an uncertainty-weighted mean. The traces can clearly be identified, as well as the 3 bright emission lines discussed in Section \ref{sec:emissionLines}. In FU Ori, the diffuse emission from the bright scattered light nebula \citep{2024A&A...686A.309Z} can be seen, as well as the trace of FU Ori S \citep{2024RNAAS...8..232C}. 

\begin{figure}[!htb]
    \centering    \includegraphics[width=0.32\linewidth]{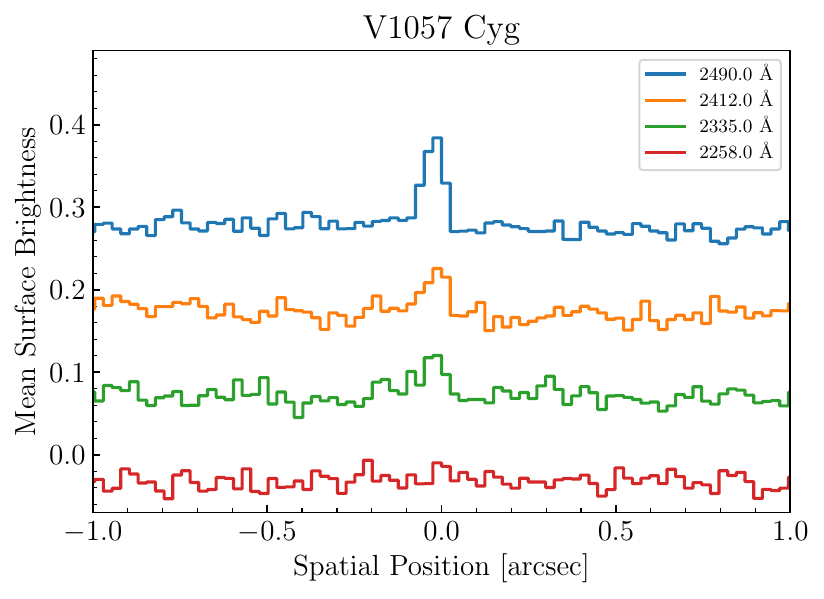}
    \includegraphics[width=0.32\linewidth]{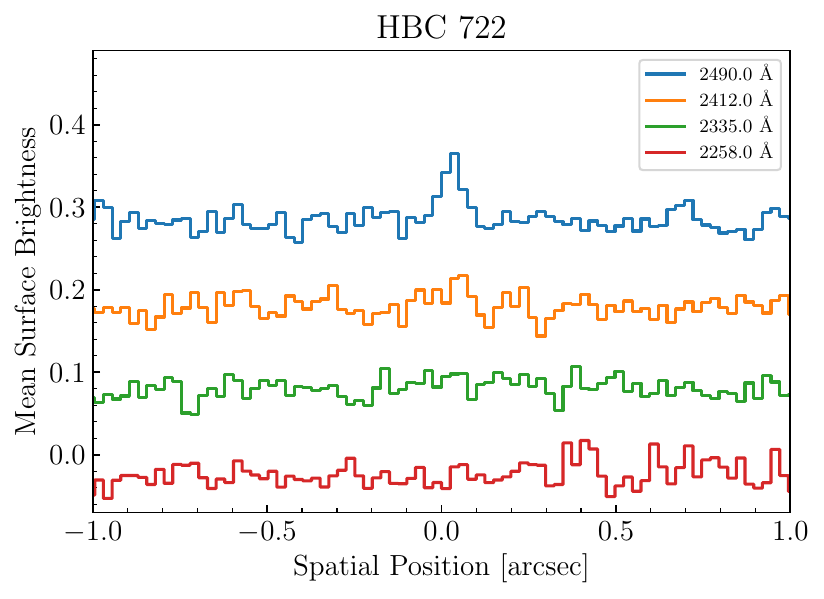}
    \includegraphics[width=0.32\linewidth]{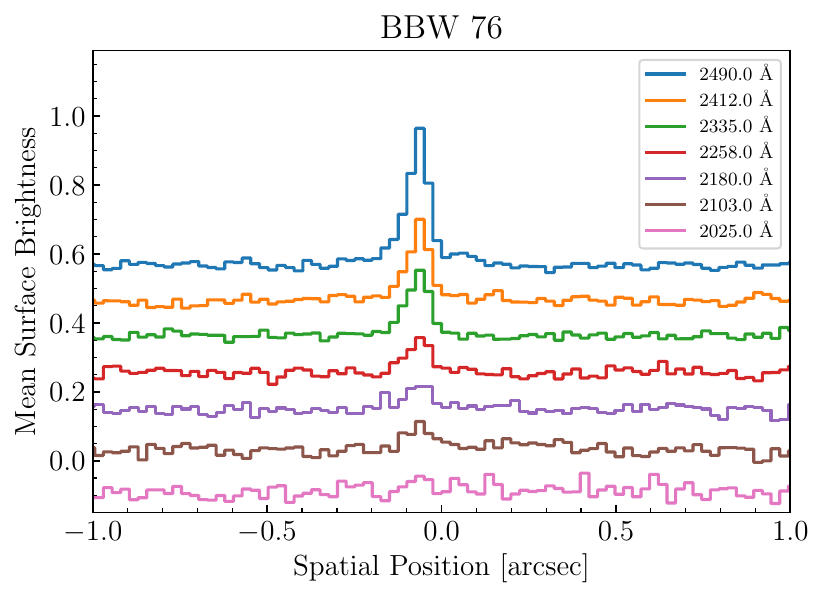}
    
    \caption{Spectrally averaged spatial cuts in the 2d spectra of V1057 Cyg (left), HBC 722 (middle) and BBW 76 (right). The spectral averaging was over $\sim 75$ \AA\ and the legend shows the mean wavelength of each bin. }
    \label{fig:detections}
\end{figure}

In order to check the wavelengths where the faintest 3 sources (V1057 Cyg, HBC 722, and BBW 76) were detected, we cut spatial profiles of the 2d spectra in $\sim78$ \AA\ bins. The spectrally-binned mean spatial profiles are shown in Figure \ref{fig:detections}. We then computed the maximum surface brightness (within $0.2^{\prime\prime}$ of the slit center) and standard deviation (within $0.2^{\prime\prime}$ of the slit center) for each spatial profile. The source was considered ``detected" in a given spectral bin if the maximum surface density was greater than $3\times$ the standard deviation. 

\begin{figure}[!htb]
    \centering    \includegraphics[width=0.45\linewidth]{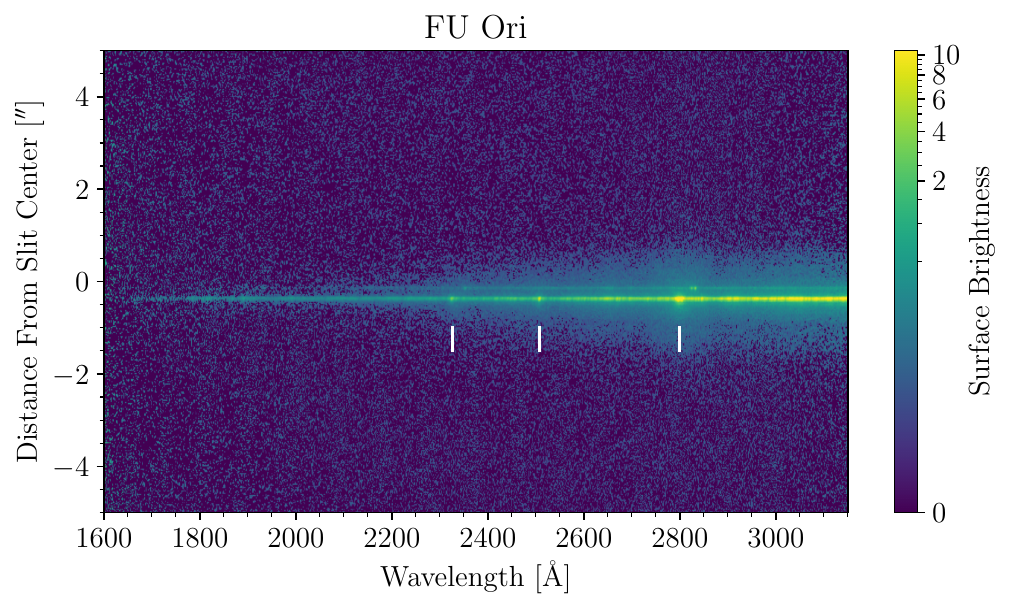}
    \includegraphics[width=0.45\linewidth]{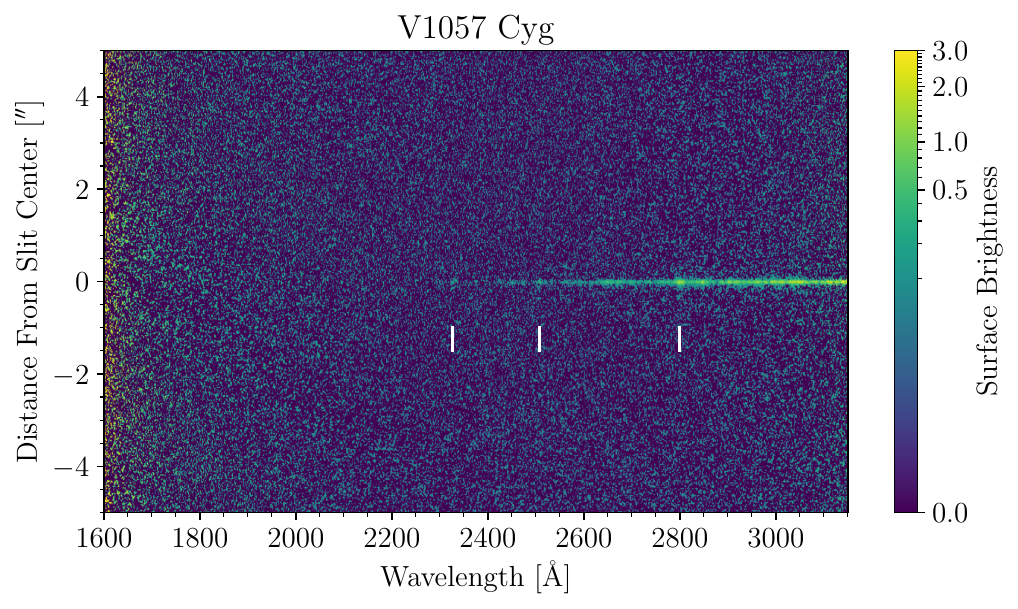}
    \includegraphics[width=0.45\linewidth]{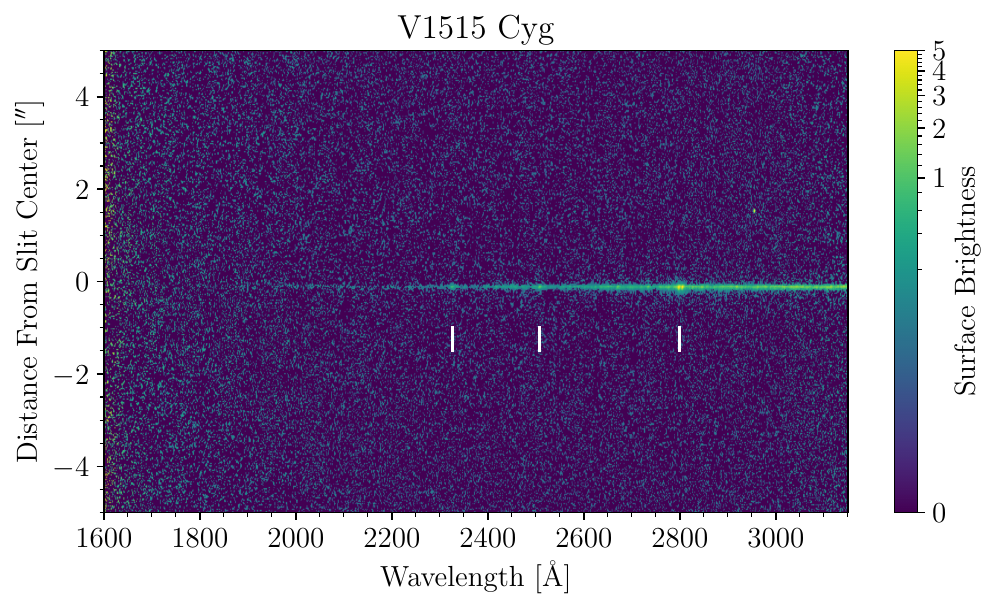}
    \includegraphics[width=0.45\linewidth]{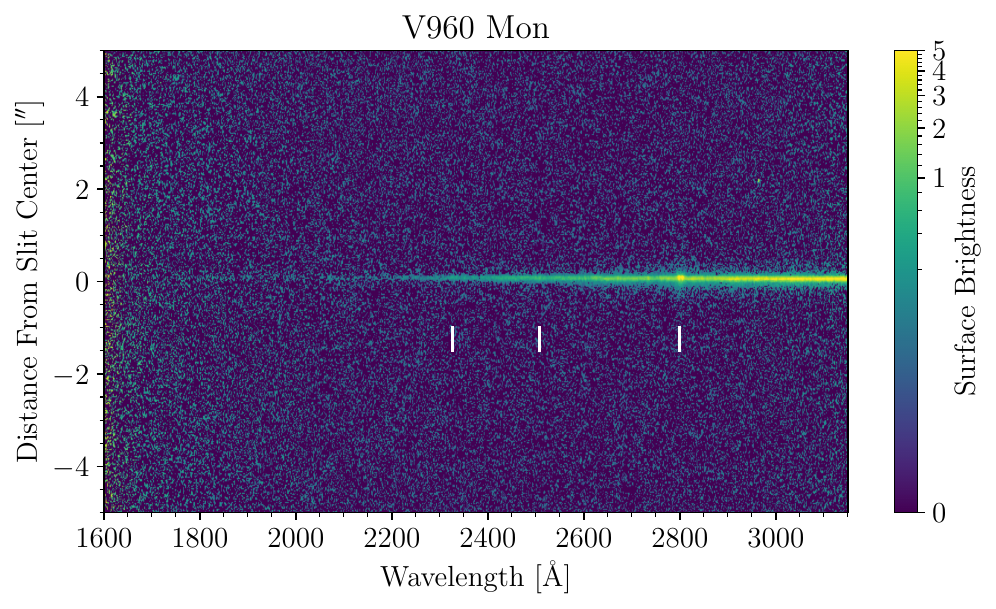}
    \includegraphics[width=0.45\linewidth]{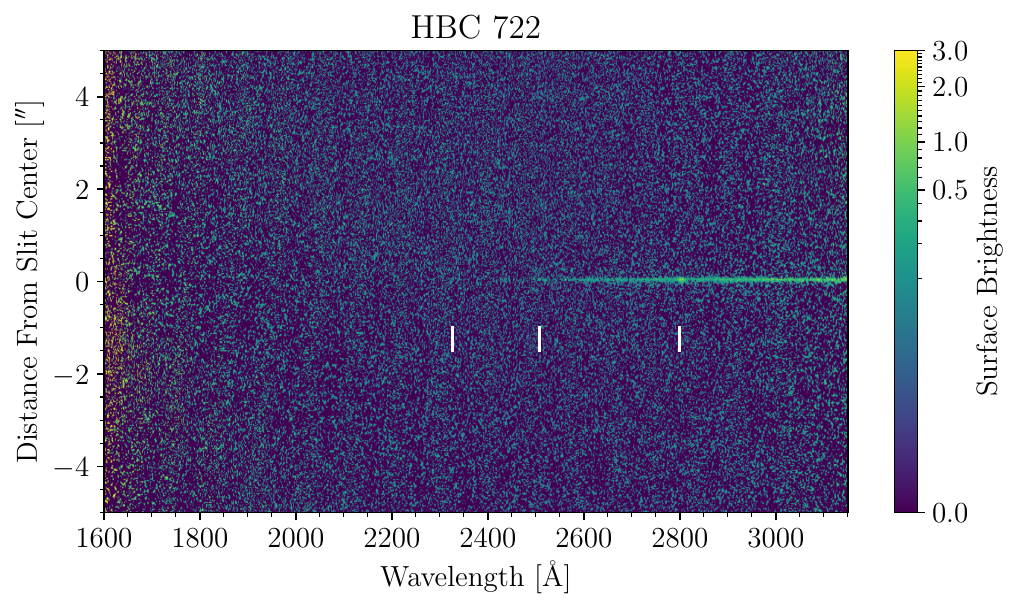}
    \includegraphics[width=0.45\linewidth]{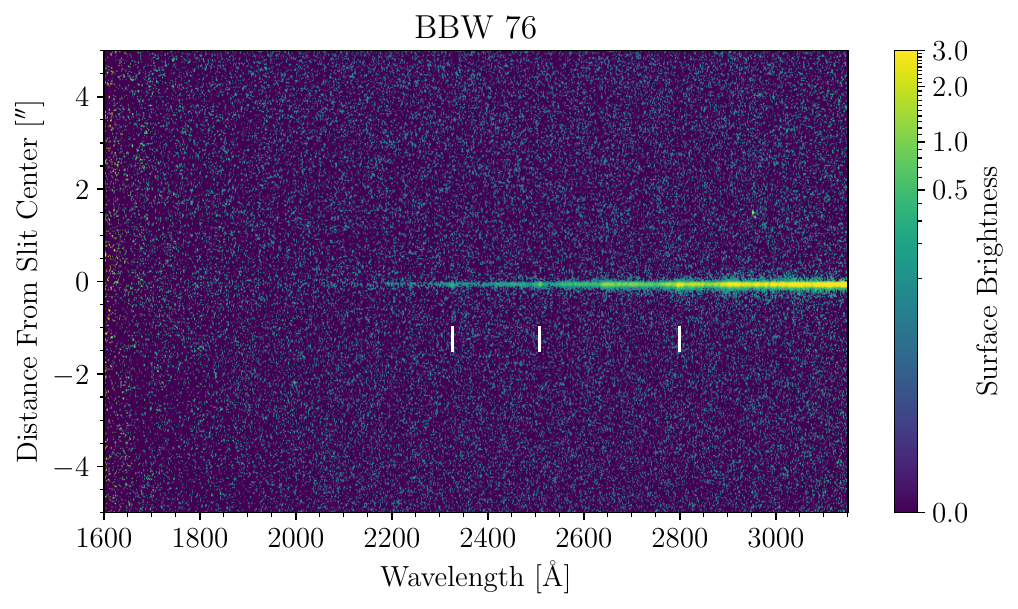}
    \caption{The two-dimensional rectified \texttt{x2d} HST/STIS images for each of our sources. The white hashes mark the locations of the \ion{C}{2}] 2326, \ion{Fe}{2}] 2500, and \ion{Mg}{2} 2800 features and the surface brightness units are $10^{-14}$ erg s$^{-1}$ cm$^{-2}$ \AA$^{-1}$ arcsec$^{-2}$. Notice that the traces of FU Ori, V1515 Cyg, and V960 Mon are all detected down to at 1800 \AA. BBW 76 is only marginally detected at 2100 \AA. V1057 Cyg is only detected redward of the \ion{C}{2} feature, with barely any continuum until 2400 \AA, and HBC 722 is only detected redward of 2450 \AA. }
    \label{fig:2dspec}
\end{figure}

\section{Posterior Distributions for the UV Excess Blackbody Fits} \label{app:CornerPlots}

\begin{figure}
    \centering
    \includegraphics[width=0.33\linewidth]{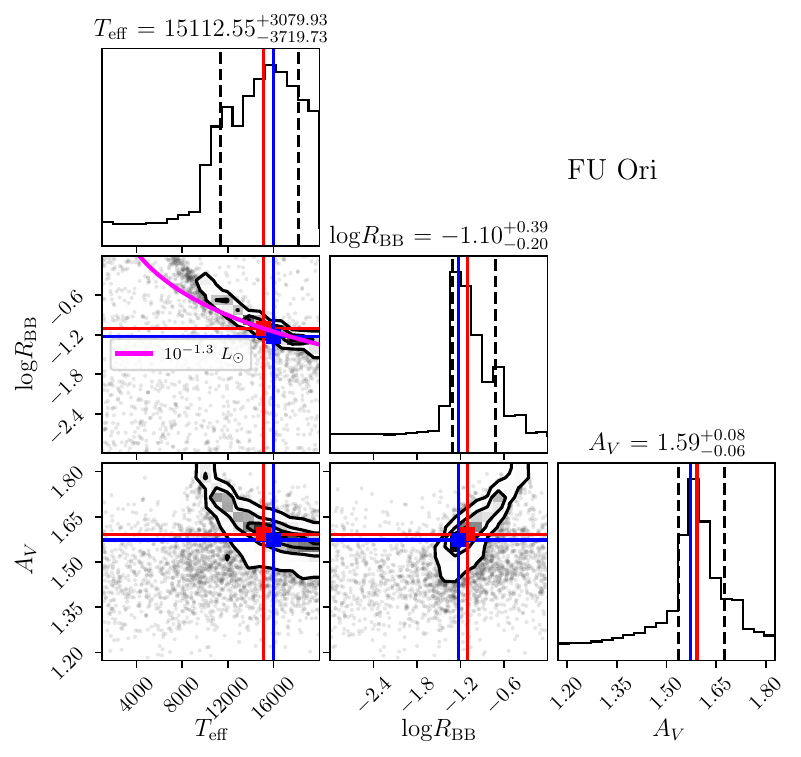}
    \includegraphics[width=0.33\linewidth]{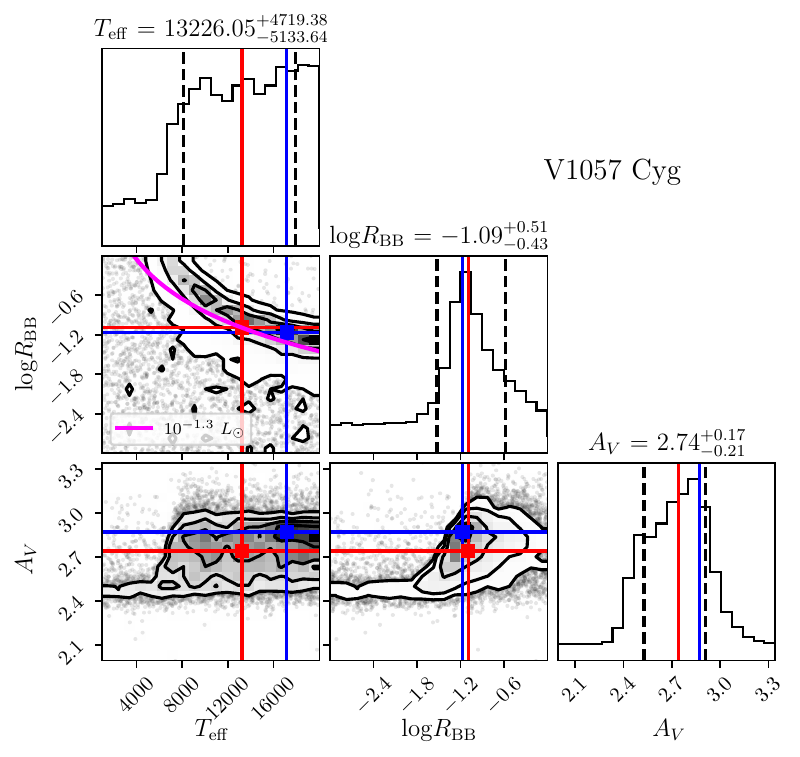}
    \includegraphics[width=0.33\linewidth]{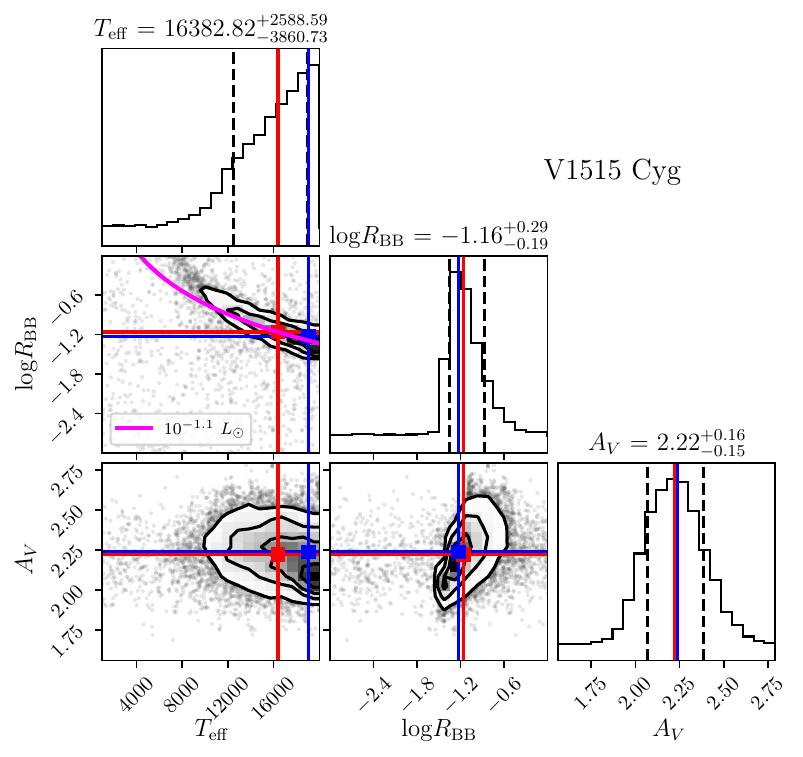}
    \includegraphics[width=0.33\linewidth]{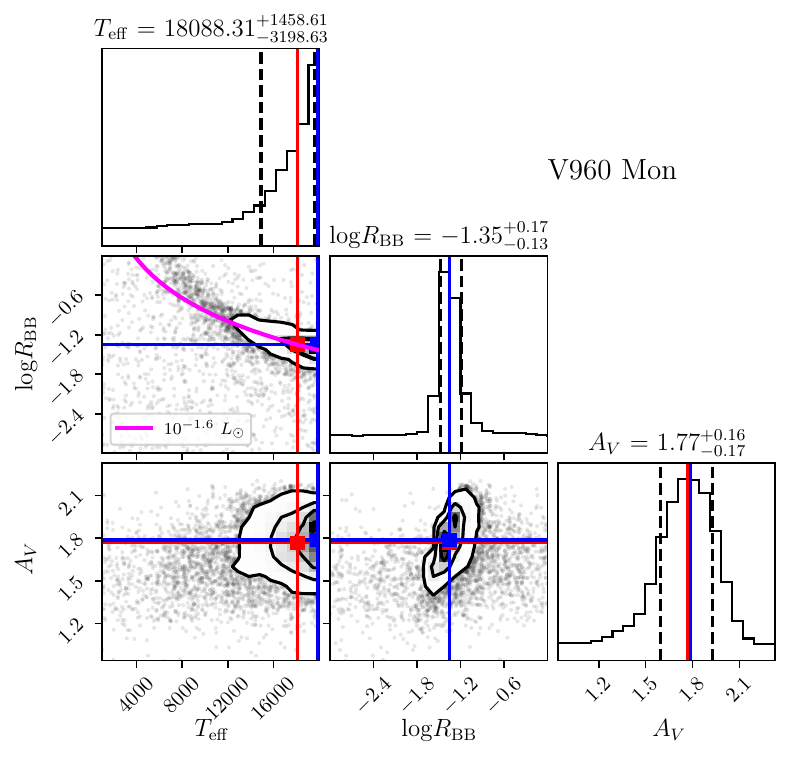}
    \includegraphics[width=0.33\linewidth]{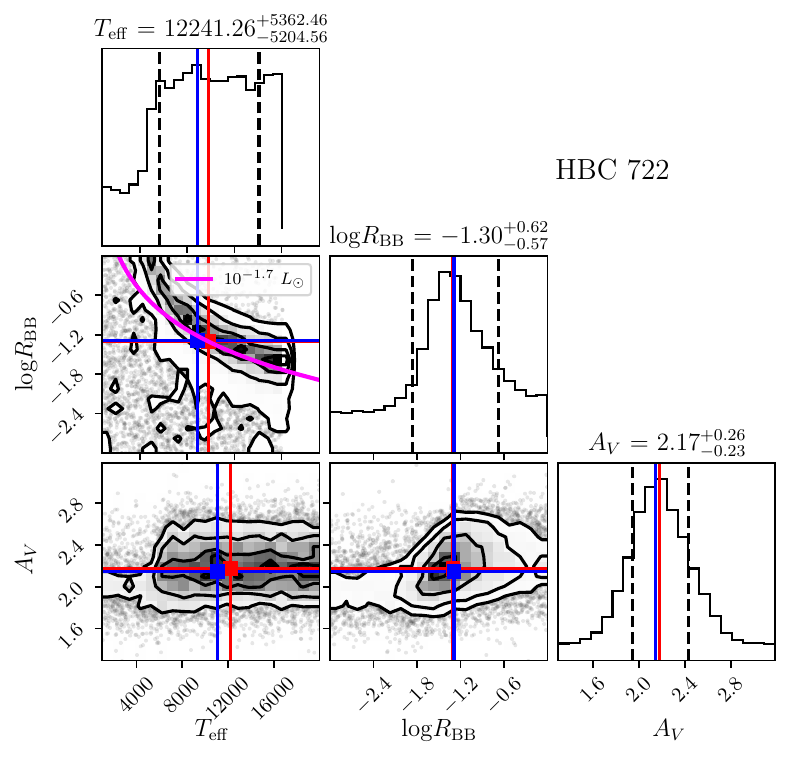}
    \includegraphics[width=0.33\linewidth]{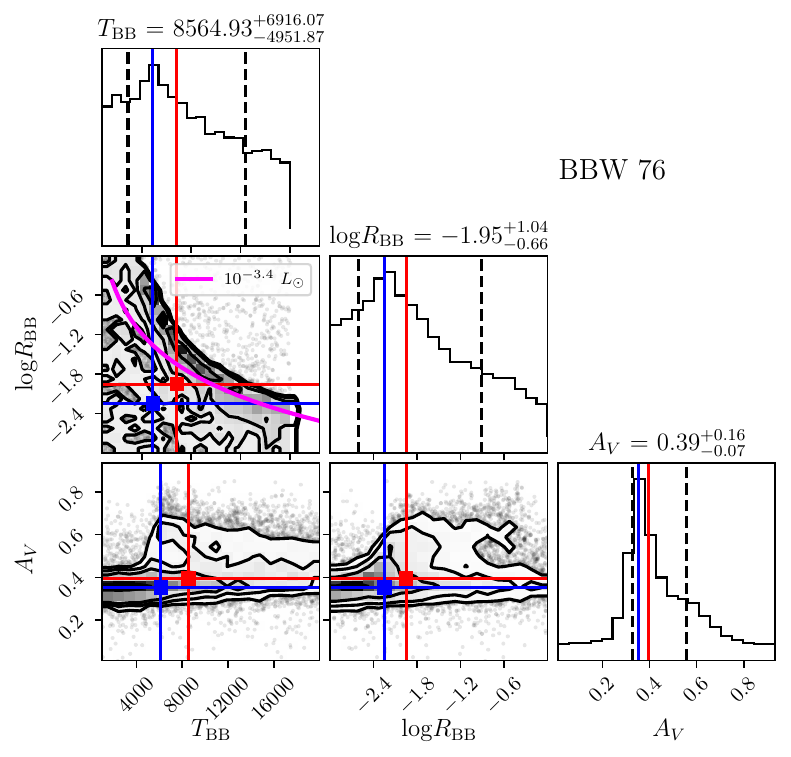}
    \caption{The posterior distributions for the blackbody fit described in Section \ref{sec:excessFit}. The red vertical and horizontal lines mark the median values for histogram, while the blue lines mark the modal values. The magenta line marks the $R_\mathrm{BB}\propto T_\mathrm{BB}^{-2}$ relation, anchored at the median $L_\mathrm{excess}$ value for each source.}
    \label{fig:CornerPlots}
\end{figure}

The posterior distributions for the blackbody fits to the UV excess are shown in Figure \ref{fig:CornerPlots} as \texttt{corner} plots \citep{corner_FM_2016}. Notice that for the sources with the strongest detected excess, the maximum temperature is poorly constrained by the NUV-only data. This is confirmed by the case of FU Ori, which has a well-constrained $T_\mathrm{BB}$ and $R_\mathrm{BB}$ in \citetalias{Carvalho_FUVFUOri_2024ApJ} when the fit included the FUV continuum but not in this paper when the fit is restricted to only the NUV. The luminosity of the emission, however, is clearly well-constrained. This can be seen from the fact that the posteriors of the $T_\mathrm{BB}$ and $R_\mathrm{BB}$ parameters lie along a locus well-described by $R_\mathrm{BB}\propto T_\mathrm{BB}^{-2}$. The $A_V$ posteriors reflect the priors, retrieving within errors the same values as the disk model best-fit values.

\section{BP Tau as a representative CTTS} \label{app:BPTau}

We select BP Tau as our reference CTTS for the comparison between the NUV spectra of FUOrs and those of non-outbursting YSOs. The $M_* \sim 0.5 \ M_\odot$ of BP Tau is comparable to those of the central objects in our sample, while its accretion rate ($\dot{M} = 10^{-7.29} \ M_\odot$ yr$^{-1}$) is typical for a CTTS \citep{Manara_PPVIIChapter_2023ASPC}. 

To determine how well the NUV spectrum of BP Tau represents typical NUV spectra of CTTSs, we query all 87 of the STIS/G230L for sources classified as T Tauri Stars in the ULLYSES archive on MAST \citep{2025ApJ...985..109R}. The stellar properties of the sample span spectral types of M6 to A2 (though most are M2-K1) and a broad range of accretion rates.

We then perform a signal-to-noise selection, keeping only the 50 spectra that have a median signal-to-noise ratio of 5 for $\lambda > 2100$ \AA. We divide each spectrum by an approximate continuum, estimated using the asymmetric least-squares fitting code described in \citep{carvalho_V960MonSpectra_2023ApJ}. We use an extremely inflexible regularization parameter of $10^12$ and fit only the 10th percentile of flux, to avoid emission features and divide out only the general slope of each spectrum. We then interpolate each spectrum to a common wavelength grid and compute the 16th, 50th, and 84th percentile normalized flux over all spectra in each wavelength bin. This enables us to estimate the median spectrum in the sample as well as the typical object-to-object variation in features like strong emission lines. 

\begin{figure}
    \centering
    \includegraphics[width=0.99\linewidth]{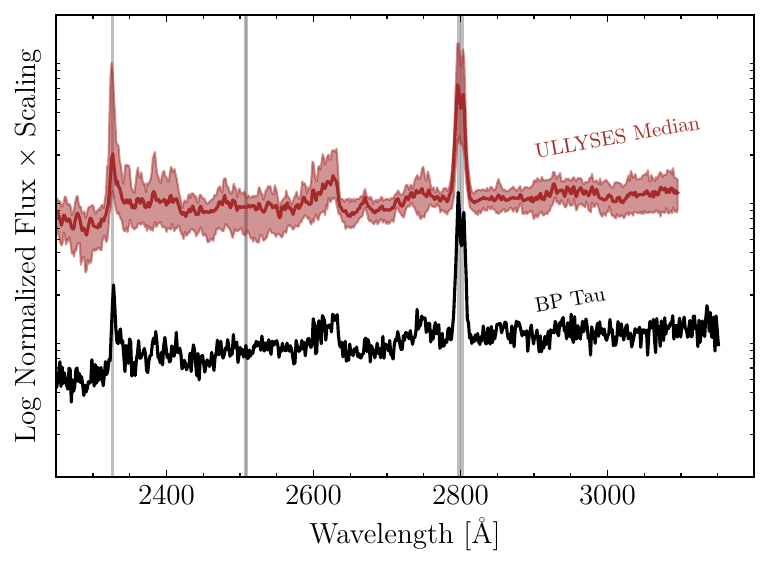}
    \caption{The G230L spectrum of BP Tau (black) compared with the median (brown line) and range (brown filled region) of G230L spectra of 50 CTTSs in the ULLYSES sample. The grey lines mark the locations of the NUV emission lines discussed in Section \ref{sec:emissionLines}. Notice that there is no sign of the \ion{Fe}{2}] emission feature at 2505 in either BP Tau or the ULLYSES sample. }
    \label{fig:BPTau}
\end{figure}

Figure \ref{fig:BPTau} shows the G230L spectrum of BP Tau compared with the median ULLYSES spectrum. The upper and lower ranges of spectra are shown in the shaded region. Notice that BP Tau almost perfectly matches the median spectrum. 

\end{document}